\documentclass[%
 reprint,
 amsmath,amssymb,
 aps,
]{revtex4-1}

\usepackage{graphicx}
\usepackage{dcolumn}
\usepackage{bm}

\newcommand{\ie}{\textit{i}.\textit{e}.}

\newcommand{\aap}{Astronomy \& Astrophysics}

\begin{document}


\title{Actinide crystallization and fission reactions in cooling white dwarf stars}

\author{C. J. Horowitz}\email{horowit@indiana.edu}
\affiliation{Center for Exploration of Energy and Matter and
                  Department of Physics, Indiana University,
                  Bloomington, IN 47405, USA}

\author{M. E. Caplan}
 \email{mecapl1@ilstu.edu}
\affiliation{
 Illinois State University, Department of Physics, Normal, IL 61790 
}%
                                    
\date{\today}

\begin{abstract}
The first solids that form as a cooling white dwarf (WD) starts to crystallize are expected to be greatly enriched in actinides.  This is because the melting points of WD matter scale as $Z^{5/3}$ and actinides have the largest charge $Z$.   We estimate that the solids may be so enriched in actinides that they could support a fission chain reaction.  This reaction could ignite carbon burning and lead to the explosion of an isolated WD in a thermonuclear supernova (SN Ia).  Our mechanism could potentially explain SN Ia with sub-Chandrasekhar ejecta masses and short delay times.




\end{abstract}

\maketitle


Phase separation and crystallization can play important roles during the cooling of white dwarf (WD) stars.  The Gaia space observatory has determined parallax distances to large numbers of galactic stars \cite{GaiaNoAuthors}, which allow for unprecedented modeling of WD and their evolution. Core crystallization, long predicted, is now resolved \cite{VanHorn1968,Tremblay2019}.  Phase separation or sedimentation of the neutron rich isotope $^{22}$Ne, in a C/O WD, could release significant gravitational energy and delay cooling \cite{Bildsten_2001,PhysRevE.82.066401,2020arXiv200713669B,Camisassa2020,Cheng_2019,bauer2020multigigayear}.

Our recent molecular dynamics (MD) simulations find that Ne has a charge that is too close to the charges of C and O for large scale phase separation upon crystallization \cite{Caplan_2020}.  However, impurities with significantly larger charges $Z$ will likely phase separate.  Material at WD densities is ionized and crystallizes to form a coulomb solid where the melting temperature scales with $Z^{5/3}$.  Pure uranium ($Z=92$) has a melting temperature 95 times higher than the melting temperature of C, and will phase separate upon crystallization.
{\it When a WD starts to crystallize, the first solids will be very strongly enriched in actinides, because they have the highest $Z$.}  This should be true even if the initial actinide abundance is very low.            

These first solids could be so enriched that they support a fission chain reaction.  For example, very high grade uranium ore deposits in Gabon, Africa became natural fission reactors 2.0 Gy ago \cite{GAUTHIERLAFAYE19964831}.  At that time, the ore had a $^{235}$U enrichment fraction  of $f_5=3.7\%$ (compared to $^{238}$U, today $f_5= 0.7\%$).  This allowed the naturally deposited ore to become critical, as first suggested by Kuroda in 1956 \cite{doi:10.1063/1.1743058}.        

A fission chain reaction in a crystallizing WD could possibly ignite carbon burning and produce a thermonuclear supernova (SN Ia).  These stellar explosions are important distance indicators in cosmology \cite{Abbott_2019,SN_cosmology,Sullivan2010} and the 2011 Nobel Prize in Physics was awarded for using SN Ia to discover the accelerating Universe and dark energy \cite{Nobel}. The exact SN Ia explosion mechanism is poorly understood but is thought to involve a WD interacting with a {\it binary} companion that is itself either a WD or a conventional star \cite{2012NewAR..56..122W,hillebrandt2013understanding,RUIZLAPUENTE201415}.  
Here we propose a completely new mechanism that involves using a fission chain reaction to ignite carbon burning in an {\it isolated} WD.

A WD first crystallizes in its high density center.  Here the density could be of order $\rho=10^8$ g/cm$^3$.  If an appropriate fuel mixture can crystallize, its critical mass may be very small.  The critical mass scales with $M_{\rm crit}\propto\lambda^3\rho$, and the neutron mean free path $\lambda$ scales with $1/\rho$ so that $M_{\rm crit}\propto 1/\rho^2$.  The critical mass could be as small as $M_c\approx10^{-6}$ g at WD densities.   Matter in a WD is very degenerate.  Therefore a fission reaction, if started, would likely be unstable.  Because of the degeneracy, a large increase in temperature need not increase the pressure very much or significantly reduce the density.  Under these conditions a fission reaction could increase the temperature enough for carbon ignition.     

In this paper we discuss some of the issues necessary to determine if a chain reaction is possible.  We start by calculating phase separation of actinides upon crystallization.  We also determine the concentration of a number of impurities that may be present.  These could absorb neutrons and prevent a chain reaction.  We make a simple estimate of the neutron multiplication factor and the minimum $^{235}$U enrichment that is necessary for criticality.  Next, we discuss the growth of a uranium rich crystal to a critical mass  and the initiation of a fission chain reaction by a neutron from spontaneous fission.
Finally, we explore if this fission reaction can ignite carbon burning and possibly lead to a thermonuclear supernova. We close with a very preliminary comparison of our fission mechanism with some SN Ia observations.

{\it Phase separation:} Consider a C/O WD made of an approximately 50/50 mixture of C and O with a very small amount of U with a mass fraction of order $10^{-10}$.  Although there has been some work on phase separation for high $Z$ elements in neutron stars, see for example \cite{Mckinven_2016}, we are not aware of applications to atomic numbers as high as 92.
As a first approximation, we simplify this three component C/O/U system by replacing the C/O mixture with a single component of average charge $Z=7$.  This should be a reasonable approximation given the similar charges of the C and O and the very large difference in charge with U.  Note that when a C/O mixture freezes, the solid is somewhat enriched in O \cite{PhysRevLett.104.231101}.  However this is a small effect.    

We construct an approximate phase diagram for the two component N/U system.   We employ the formalism of Medin and Cumming that assumes the free energy of the system follows from linear mixing rules plus small corrections \cite{PhysRevE.81.036107}.  We caution that the Medin and Cumming free energy fits may have larger errors for the very large ratio of charges 92/7 that we consider.  Thus our phase diagram may be somewhat preliminary.  

Figure \ref{Fig1} shows our phase diagram.  
The system starts in the liquid phase with a very small U number fraction of order $x_2\approx 10^{-11}$.   The melting temperature of pure U is $73 T_1$.  However, the very large entropy of mixing keeps the tiny U fraction dissolved in the liquid until much lower temperatures.  Finally, at $T\approx 1.8T_1$ the large lattice energy overcomes the entropy of mixing and the U precipitates out to form a solid that is greatly enriched in U.  
The original U number fraction $x_2$ is somewhat uncertain.  Actinide boost stars, that are relatively enriched in actinides, have now been observed \cite{Holmbeck_2018}.   However, this may not be so important.  We emphasize that {\it the solid in Fig. \ref{Fig1} is greatly enriched in U even if the initial $x_2$ is very small}.   

\begin{figure}[tb]
\centering  
\includegraphics[width=0.5\textwidth]{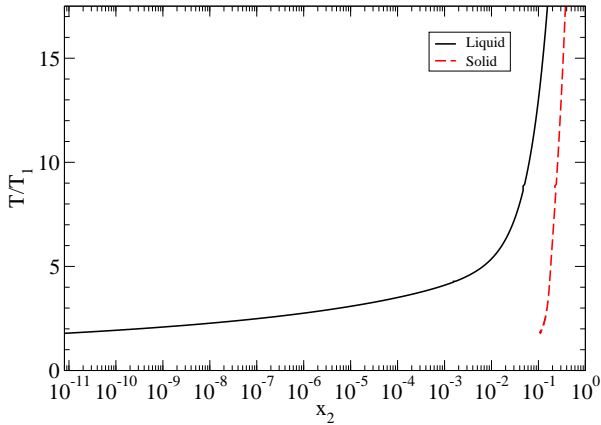}
\caption{\label{Fig1} Phase diagram for a mixture of N and U.  The y axis is the temperature in units of the melting temperature of pure N (about 0.5 keV) and the x axis is the number fraction of U.  The liquid phase, with composition given by the solid black line, is in equilibrium with the solid phase (dashed red line).}	
\end{figure}

We now consider a range of additional impurities that might absorb $n$.  The solar abundance of Pb is $\approx$100 times that of the actinides.  The charge of Pb is close enough to that of U so that some Pb will likely be present in the actinide rich solid.  

To explore impurity concentrations, we construct a multicomponent phase diagram.  We assume a linear mixing rule for the free energies of multicomponent systems.  This corresponds to the formalism of Medin and Cumming with the further approximation that the term describing corrections to linear mixing $\Delta f_s$ is set to zero.  This approximation has minimal effect on the two component phase diagram in Fig. \ref{Fig1}.

The abundance, by number, of species i in the liquid $a_i$ is related to the abundance in the solid $b_i$ by \cite{PhysRevE.81.036107},
\begin{equation}
\delta f^{\rm OCP}(\Gamma_i)+ {\rm ln}\frac{a_i Z_i}{\langle Z\rangle_a}-\frac{Z_i}{\langle Z\rangle_a}={\rm ln}\frac{b_i Z_i}{\langle Z\rangle_b}-\frac{Z_i}{\langle Z\rangle_b}\, .
\label{eq.df}
\end{equation}
Here $Z_i$ is the atomic number of species $i$,  $\langle Z\rangle_a=\sum_i a_i Z_i$ and $\langle Z\rangle_b=\sum_i b_i Z_i$. The difference in free energy of the liquid and solid phases of a one component plasma is $\delta f^{\rm OCP}(\Gamma_i)$.  
We use the expression in Eq.~9 of Medin and Cumming for $\delta f^{\rm OCP}$ \cite{PhysRevE.81.036107}.  However, we caution that this involves poorly known free energies for very super cooled liquids and super heated solids.  The free energy difference $\delta f^{\rm OCP}$  is for a coulomb parameter $\Gamma_i$ that describes the ratio of coulomb to thermal energies.  The coulomb parameter for species $i$ is related to $\Gamma_1$ of species one by $\Gamma_i=(Z_i/Z_1)^{5/3}\Gamma_1$ and each $\Gamma_i$ is proportional to one over the temperature.  We adjust the temperature until $\sum b_i=1$.

We start with the solar abundances (at early times) from ref. \cite{lodders2019solar} and assume that all of the original C, N, and O has been converted into $^{22}$Ne.  We also assume all of the elements with Z=1-5 are converted into equal numbers of $^{12}$C and $^{16}$O.  This gives the liquid phase abundances shown in Fig. \ref{Fig2} along with solid abundances from Eq. \ref{eq.df}.  These abundances are tabulated in the supplemental material \cite{supplemental}. 

\begin{figure}[htb]
\centering  
\includegraphics[width=0.52\textwidth]{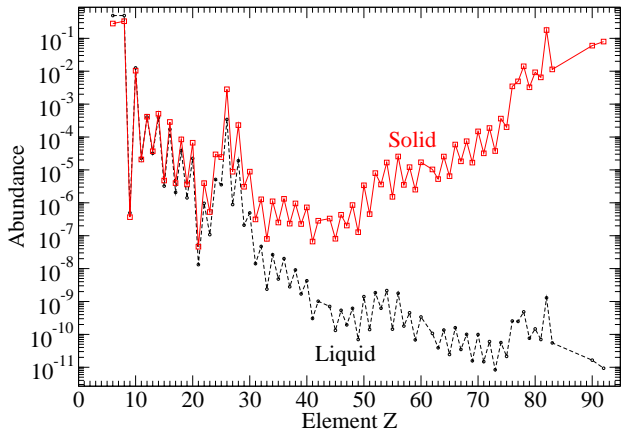}
\caption{\label{Fig2} Composition of the equilibrium solid phase (shown by the full red line) and liquid phase (dashed black line) for all chemical elements $Z$.  This assumes a linear mixing approximation for the free energies, see text. Numerical values are listed in the supplemental information \cite{supplemental}.}	
\end{figure}

To refine these compositions and check the linear mixing approximation, we have performed molecular dynamics (MD) simulations as shown in the supplemental materials \cite{supplemental}.  These find that the abundance of C+O in the solid could be less than or of order 40\%.  As a result we assume the solid has the composition given in Tab.~\ref{Table1} which is 40\% C+O and 60\% Pb+Th+U where the relative concentrations of Pb, Th, and U are taken from the results of Eq. \ref{eq.df}.


\begin{table}[tb]
\caption{\label{Table1} Composition of solid (abundance by number).}
\begin{tabular*}{0.2\textwidth}{c c c c c } 
C & O & Pb  & Th & U \\ \hline
 0.20& 0.20 & 0.32 & 0.12 & 0.16 \\
\end{tabular*}
\end{table}

{\it Criticality:} For the composition in Tab.~\ref{Table1}, we calculate
the multiplication factor which is equal to the number of fissions in one generation over the fissions in the proceeding generation, 
\begin{equation}
k_\infty=\frac{\nu f_5 \sigma_f(^{235}{\rm U})}{f_5\sigma_a(^{235}{\rm U})+(1-f_5)\sigma_a(^{238}{\rm U})+N_{\rm Th} \sigma_a(^{232}{\rm Th})}\, .
\label{eq.k}
\end{equation}
Here $\sigma_f(^{235}{\rm U})$ is the fission cross section for $^{235}$U and $\sigma_a(^{235}{\rm U})$ is the sum of $\sigma_f(^{235}{\rm U})$ and $\sigma_{n,\gamma}$ for the $n,\gamma$ reaction.  Likewise $\sigma_a(^{238}$U) and $\sigma_a(^{232}$Th) are the $n,\gamma$ cross sections for $^{238}$U and $^{232}$Th.  Finally $\nu$ is the number of $n$ emitted per fission and $N_{\rm Th}$ is the number density of Th over the number density of U.  We assume $N_{\rm Th}=0.75$ from Tab.~\ref{Table1}.

We start by evaluating all cross sections at a single energy $E$.  We use the ENDF data set from Dec. 2011 that is available from the National Nuclear Data Center \cite{NNDC}.  In Tab. \ref{Table2} we present the minimum  $^{235}$U enrichment $f_5$ so that $k_\infty\ge 1$.  In general this enrichment increases slightly as $E$ decreases.

\begin{table}[tb]
\caption{\label{Table2} Minimum $^{235}$U enrichment $f_5$ necessary for criticality assuming mono-energetic neutrons of energy $E$.}
\begin{tabular*}{0.33\textwidth}{c c c c c c c } 
$E$(MeV) & 1.0 & 0.5  & 0.25 & 0.125& 0.1&0.05 \\ \hline
 $f_5$& 0.12 & 0.14 & 0.14 & 0.16 & 0.18 & 0.26\\
\end{tabular*}
\end{table}

Neutrons will lose energy scattering from C and O.  After $n_{\rm col}$ collisions the initial energy $E_0$ will be reduced to $E=E_0e^{-\bar\xi n_{\rm col}}$. Here $\bar\xi=1-\frac{(A-1)^2}{2A}\ln\big(\frac{A+1}{A-1}\big)\approx 0.139$ for a nucleus of mass number $A$ and we have taken the average of C and O \cite{Lamarsh}.  The number of collisions before a $n$ is absorbed depends on scattering and absorption cross sections and the ratio of the number densities of C+O to U.
We estimate $n_{col}\approx 15$, although this number will be smaller if the crystal contains less C and O than assumed in Tab.~\ref{Table1}.  If $E_0\approx 1$ MeV (the average energy of the initial fission spectrum is 2 MeV) than $E\approx 0.12$ MeV and Tab.~\ref{Table2} suggests the system will be critical for $f_5\ge 0.16$.  We emphasize that this is a first estimate, that also depends on uncertain compositions, and should be verified in future work.  
Note that we have neglected $n$ absorption on Pb.  There are some resonances in Pb isotopes with energies below 0.5 MeV that will absorb a few $n$.   This should also be studied further. 

{\it Delay time:} Before the actinides can crystallize, there will be a total delay time consisting of the main sequence lifetime of the original star plus the cooling time of the WD.  During this time $^{235}$U and $^{238}$U will decay with half lives of 0.7 and 4.5 Gy respectively.  For low mass WD this delay time can be very long and the remaining $^{235}$U fraction will be low.   Therefore, we focus on WD with masses above $\approx M_\odot$.  These stars have much shorter delay times.  For example a $1.1M_\odot$ WD will start to crystalize after cooling for about 1 Gy \cite{bauer2020multigigayear}. 
According to Fig.~\ref{Fig1}, the U precipitates out at a temperature of about 1.8 times the melting temperature $T_1$ of the background, thus requiring much less than 1 Gy of cooling time. From the initial-final mass function, a $1.1M_\odot$ WD formed from main sequence star of perhaps $5-6M_\odot$ with a main sequence lifetime of less than 0.2 Gy \cite{Cummings_2018}. Therefore, the total delay time can be significantly less than 1 Gy for a massive WD.

We don't know the original enrichment (isotopic fraction) of $^{235}$U, $f_{5}(t=0)$.  This could depend on the time between nucleosynthesis events and star formation.  As an example, we assume $f_{5}(0)=0.25$.  
This is the early solar system value, \ie\ $f_{5}(0)=0.25$ when the solar system formed so that radioactive decay leads to the present day value $f_{5}$(4.6~Gy)=0.007 \cite{lodders2019solar}.   Given $f_5(0)=0.25$, $f_5(t=0.7$~Gy$)=0.16$.  Therefore, if $f_5\ge 0.16$ is needed for the system to be critical, the delay time will need to be less than 0.7 Gy. We show $f_5(t)$ in Tab.~\ref{Table3} and this can be compared to the necessary critical values in Tab.~\ref{Table2} to determine the maximum allowed delay time.

\begin{table}[tb]
\caption{\label{Table3} Enrichment fraction of $^{235}$U $f_5(t)$ versus delay time $t$.}
\begin{tabular*}{0.35\textwidth}{c c c c c c c } 
$t$(Gy) & 0 & 0.2  & 0.4 & 0.6 & 0.8 & 1.0 \\ \hline
 $f_5(t)$& 0.250 & 0.220 & 0.193 & 0.168 & 0.146 & 0.126\\
\end{tabular*}
\end{table}

{\it Critical mass assembly and fission ignition:} 
It is natural to think that as the crystal approaches the critical mass there will be a steady increase in fission heating, which could burn off the U fuel or melt the crystal. We show below that this is not the case, and that diffusion-driven growth of the crystal proceeds very rapidly and grows well in excess of a critical mass before a $n$ initiates a chain reaction.  

The solid can grow as U diffuses to a small seed crystal. Consider a spherical volume of liquid that contains (in very diluted form) one critical mass $M_c\approx 10^{-6}$ g of U.  This will have a total mass $M_{\rm tot}=M_c/x_U\approx 10^4$ g, given that the mass fraction of U is $x_U\approx 10^{-10}$.  At a density $\rho=10^8$ g/cm$^3$, the sphere has a radius $r=[3M_{\rm tot}/(4\pi\rho)]^{1/3}=0.029$ cm.   For the crystal to grow to $M_c$, U will need to diffuse over this distance $r$ which takes a time $t_D\approx r^2/D$.  Here $D$ is the diffusion constant for trace amounts of U in a C/O liquid.  At a temperature $kT\approx 3$ keV we evaluate $D$ from Eqs. 8 and 11 of ref. \cite{bauer2020multigigayear} to be $D\approx 3.6\times 10^{-5}$ cm$^2$/s giving $t_D\approx 23$ s.  

Cooper and Bildsten explore nucleation of seed crystals in the one component plasma \cite{PhysRevE.77.056405}. Nucleation in our low heavy element concentration system  should be studied further.  In addition to diffuse growth of a single crystal, multiple small crystals could assemble.  This may somewhat change our estimate for $t_D$.  However as we discuss below, $t_D$ is fully six orders of magnitude shorter than the time for a neutron from spontaneous fission to initiate a chain reaction.  Therefore, we do not expect the assembly of multiple crystals to change our main results.

 We compare $t_D$ to the time for $n$ emission that might start (or perhaps prematurely start) a chain reaction.  The $n$ background is very likely dominated by local sources because the system is self shielding.  The density is so high that $n$ can't diffuse in from the outside.  The local $n$ background is probably from spontaneous fission of $^{238}$U with a partial half-life of $8.4\times 10^{15}$ y.  Alpha decay followed by $^{13}$C$(\alpha,n)$ or $^{22}$Ne$(\alpha,n)$ is not a problem because the $\alpha$ will quickly thermalize in the ionized plasma and then the thermal $(\alpha,n)$ rate is very low.  Spontanious fission of transuranic elements could possibly contribute.  However, a relatively long half-life and high initial abundance might be needed to survive the delay time.  The total spontaneous fission rate of $^{238}$U in $M_c=10^{-6}$ g is only $\approx 0.3$ per year.  Thus the time $t_f$ for a single emitted $n$ is of order a year.  We emphasize that this time $t_f\approx 1$ y is dramatically longer than the time $t_D\approx 23$ s for the crystal to grow to a critical mass.  We conclude that the crystal will very likely grow much larger than a critical mass before any fissions occur.
 
 We assume that the crystal will continue to grow until the diffusion time is equal to the time for one spontaneous fission to occur in the crystal.  This yields a crystal mass of $M^*\approx 5$ mg after a time of $\approx 1.8$ h.  This mass is highly supercritical, $M^*\approx 5000 M_c$. 
 
 When a chain reaction is finally initiated, the reaction will progress extremely rapidly and release a total energy,
 \begin{equation}
 E_{\rm tot} \approx 200 {\rm \ MeV} n_U\epsilon \ \ \approx 4\times 10^{15} \epsilon \ {\rm ergs}\, .
 \end{equation}
 Here $n_U$ is the number of U ions in $M^*$ and $\epsilon$ is the fraction of the ions that fission.  We expect $\epsilon$ to be high because the system is highly supercritical (and thus will remain critical even if partially disassembled by the explosion). Furthermore the system is degenerate so the temperature can rise significantly before the pressure greatly increases.  This will delay the disassembly of the system.  
 
 The large energy release will greatly increase the temperature.  As long as the electrons are degenerate, the heat capacity will be dominated by the ions $C\approx \frac{3}{2}(1+N_{\rm Th}+N_{\rm Pb} +N_{\rm C+O})n_U$.  Here $N_{\rm Pb}$ is the number density of Pb over the number density of U etc.  Using the composition from Tab.~\ref{Table1} yields $C=15n_U$ and a final temperature $kT\approx E_{\rm tot}/C\approx 13\epsilon$~MeV.  In reality the electron Fermi energy is only $E_F=1.9$~MeV.  We conclude the final temperature will be larger than, or of order, the electron Fermi energy $T\ge 2.2\times 10^{10}$~k.  Note that some heat will be lost via conduction because the thermal conductivity of the degenerate electrons is large, see for example \cite{PhysRevD.74.043004}.  However the fission chain reaction will proceed very rapidly with an exponentially rising energy production rate. This large rate could limit the time for heat conduction.      

{\it Carbon ignition:} Timmes and Woosley have explored the conditions necessary to ignite C burning via a deflagration \cite{1992ApJ...396..649T}.  According to their Fig.~6, a trigger mass of $M^*=5$ mg needs to be heated above $5\times 10^9$ K for carbon ignition.  Our temperature $T\ge 2\times 10^{10}$ K meets this condition.  After carbon ignition, the deflagration could possibly turn into a detonation \cite{Poludnenkoeaau7365}.  We conclude, it is plausible that a fission chain reaction could ignite a thermonuclear supernova (SN). 

We emphasize that this conclusion needs to be verified with detailed astrophysical simulations.  These simulations can explore many open issues including the important role of heat conduction losses.  In addition, future molecular dynamics simulations can improve our knowledge of the phase diagram and the amount of C and O in the crystal (which is presently uncertain).  This is important for the neutron spectrum which also impacts the fission reaction time scale and the time for heat conduction losses. 


{\it SN Ia:} A fission chain reaction initiating a SN provides a new mechanism that could explain a subset of the observed SN Ia.  Our mechanism could work for a single {\it isolated} WD and does not require either a main sequence or a degenerate companion.  
Mannucci et al. argue for a bimodal delay time distribution for observed SN Ia with about 50\% `prompt' SNe Ia with short delay times of order $10^8$ y and the remaining 50\% `tardy' SNe Ia with a much broader delay time distribution \cite{10.1111/j.1365-2966.2006.10501.x,10.1111/j.1365-2966.2006.10848.x}, however see Section 3.5 of \cite{doi:10.1146/annurev-astro-082812-141031}.  Our mechanism could explain a fraction of SN with relatively short delay times.   We assume SN with delay times longer than the half-life of $^{235}$U come from a more conventional mechanism involving one or two WDs in a binary system (\cite{2012NewAR..56..122W,hillebrandt2013understanding,RUIZLAPUENTE201415}).



{\it Conclusions:} The first solids that form as a white dwarf (WD) starts to crystallize are greatly enriched in actinides because of their large charges.  We estimate that these first solids could be so enriched in actinides that they may support a fission chain reaction.  This reaction could ignite carbon burning and lead to the explosion of an isolated WD in a thermonuclear supernova.

We thank Andrew Cumming, Gerardo Ortiz, Irina Sagert, and Mike Snow for helpful discussions.  This research was supported in part by the US Department of Energy Office of Science grants DE-FG02-87ER40365 and DE-SC0018083.

\providecommand{\noopsort}[1]{}\providecommand{\singleletter}[1]{#1}%

\section*{Appendix}
\subsection{Linear mixing compositions}
\label{Sec.phase}

In this section we provide in Tables \ref{Table1} and \ref{Table2} abundances of the liquid and solid phases as shown in Fig. 2 of the main paper.  Molecular dynamics simulations are described in Sec. \ref{Sec.MD}.

\begin{table}[h]
\caption{\label{Table1} Abundance of liquid $a_i$ and solid $b_i$ phases by number (continued on next Table).}
\begin{tabular*}{.45\textwidth}{c c c } \hline \hline
$Z$ & $a_i$ (liquid) & $b_i$ (solid) \\ \hline
   6   &    0.49295194553376287   &    0.23362533854669934 \\    
   8   &    0.49295194553376287   &    0.32492889695169003 \\    
   9   &     4.9686426256180860E-007 &  3.6866819993714355E-007\\
   10  &      1.2636784000587731E-002 &  1.0409983645970053E-002 \\
   11  &      2.2613192422104362E-005 &  2.0796337324649017E-005 \\
   12  &      4.0296865388871094E-004 &  4.1472075483349973E-004\\
   13   &     3.2010577923470222E-005 &  3.6915127207336077E-005\\
   14  &      3.9123170280457373E-004 &  5.0828508971819709E-004\\
   15  &     3.2315738651657788E-006  & 4.7577577604337754E-006\\
   16  &      1.7096825412559870E-004 &  2.8688838579082860E-004\\
   17  &      2.0696157078361951E-006 &  3.9805002804735190E-006\\
   18   &     3.8223337364006851E-005 &  8.4726513693123020E-005\\
   19   &     1.4127376788273156E-006 &  3.6286258879588543E-006\\
   20  &     2.2391755278316973E-005 &  6.6998323717337873E-005\\
   21   &     1.3184508384514136E-008 &  4.6195637709868454E-008\\
   22    &    9.6203875719644679E-007  & 3.9675311376132979E-006\\
   23  &      1.0758871827125777E-007 &  5.2490469470911812E-007\\
   24  &      5.1368722578240533E-006  & 2.9796231189119231E-005\\
   25  &      3.5562961784935750E-006  & 2.4645733986116784E-005\\
   26 &       3.4115404484558827E-004 &  2.8384366779132123E-003\\
   27 &       8.8418364833833664E-007 &  8.8742625202445955E-006\\
   28  &      1.9041246975498605E-005 &  2.3163094474264208E-004\\
   29 &       2.0930896100044694E-007 &  3.1004590066959337E-006\\
   30 &       4.9295194553376294E-007 &  8.9326795429250368E-006\\
   31 &       1.4162587641525570E-008 &  3.1538300926850089E-007\\
   32 &       4.6947804336548850E-008 &  1.2905875582566026E-006\\
   33&        2.3747764360237624E-009&   8.0948499116693123E-008\\
   34&       2.6447263109589182E-008&   1.1227893608345454E-006\\
   35&       4.8121499444962572E-009&   2.5555642110274423E-007\\
   36&        2.0070186353874633E-008&   1.3390891593029508E-006\\
   37&       2.8051313091087936E-009&   2.3614812413699321E-007\\
   38&        9.1156986753465687E-009&   9.7238962176758543E-007\\
   39&      1.7018579071998956E-009&   2.3100543426260367E-007\\
   40&        4.2644255605698538E-009&   7.3964223875862461E-007\\
   41&        3.0516072818756751E-010&   6.7912630607248444E-008\\
   42&        1.0172024272918918E-009&   2.9165792081199557E-007\\
   44&       7.0812938207627849E-010&   3.4116346103081000E-007\\
   45&        1.3223631554794592E-010&   8.3083051543101197E-008\\
   \hline\hline
   \end{tabular*}
   \end{table}
   
\begin{table}[h]
\caption{ \label{Table2} Abundance of liquid $a_i$ and solid $b_i$ phases by number continued.}
\begin{tabular*}{.45\textwidth}{c c c } \hline \hline
$Z$ & $a_i$ (liquid) & $b_i$ (solid) \\ \hline
   46&        5.3989974987031172E-010&   4.4413587486582836E-007\\
   47&        1.9444215629387313E-010&   2.1025715753936566E-007\\
   48&        6.1814609043122656E-010&   8.8209113811041236E-007\\
   49&       7.0030474802018688E-011&   1.3239262544836350E-007\\
   50&        1.4045218130684197E-009&   3.5313472899560926E-006\\
   51&       1.4045218130684195E-010&   4.7146037001072241E-007\\
   52&       1.8466136372375879E-009&   8.3072267398741265E-006\\
   53&      6.2205840745927222E-010&   3.7646306118210063E-006\\
   54&        2.1517743654251553E-009&   1.7584763866454542E-005\\
   55&        1.4397326663208312E-010&   1.5947680709495643E-006\\
   56&       1.7801042477608103E-009&   2.6825900187282332E-005\\
   57&      1.7957535158729935E-010&   3.6953572845768090E-006\\
   58&     4.5382877525330550E-010 &  1.2799728815374261E-005\\
   59&      6.8465547990800394E-011&   2.6562491509055975E-006\\
   60&       3.3802419122315170E-010 &  1.8105578444288122E-005\\
   62&       1.0680625486564863E-010  & 1.1023049628428485E-005\\
   63&       3.9123170280457374E-011 &  5.6350086615352327E-006\\
   64&        1.3536616917038251E-010&   2.7306869014457609E-005\\
   65&     2.4451981425285858E-011&   6.9328962431122075E-006\\
   66&      1.5923130304146148E-010&   6.3679214189768757E-005\\
   67&       3.4858744719887521E-011&   1.9732020393274201E-005\\
   68&       1.0015531591797088E-010 &  8.0526483479513336E-005\\
   69&        1.5766637623024322E-011 &  1.8068230574103112E-005\\
   70&       9.8590389106752573E-011   &1.6159298517913045E-004\\
   71&       1.4866804706573801E-011&   3.4971104242230517E-005\\
   72&      6.0640913934708923E-011  & 2.0542183007793289E-004\\
   73&      8.4114816102983345E-012   &4.1173800326116089E-005\\
   74&   5.6337365203858610E-011&   3.9983817265815087E-004\\
   75&   2.1400374143410181E-011 &  2.2095984970340151E-004\\
   76&     2.5508307022858210E-010&   3.8444691668953371E-003\\
   77&      2.4764966787529519E-010&   5.4664740070573484E-003\\
   78&     4.8512731147767138E-010 &  1.5735689736341808E-002\\
   79&      7.6290182046891878E-011&   3.6483689993806979E-003\\
   80&      1.4710312025451973E-010&   1.0406039315687302E-002\\
   81&      7.0030474802018688E-011&   7.3521024108057099E-003\\
   82&      1.2949769362831390E-009&  0.20242723306727989     \\
   83&     5.5163670095444893E-011 &  1.2881256024068765E-002\\
   90&       1.6470854688072555E-011&   6.9868308321281880E-002\\
   92&   9.3465253800012668E-012&   9.3418583343566702E-002\\
\hline \hline
\end{tabular*}
\end{table}

\eject
\subsection{Molecular dynamics simulations}
\label{Sec.MD}
In this section we describe preliminary molecular dynamics simulations that are illustrated in Fig. 3.   We have extensively used classical molecular dynamics simulations to model astromaterials in WDs and neutron stars \cite{CaplanRMP}. In our model nuclei are fully ionized and treated as point particles of charge $Z_i$ which interact via a screened two-body potential 

\begin{equation}
V(r_{ij})=\frac{Z_i Z_j e^2}{r_{ij}} \exp(-r_{ij}/\lambda).
\label{eq.V}
\end{equation} 

\noindent with periodic separation $r_{ij}$ and screening length $\lambda^{-1}=2\alpha^{1/2}k_F/\pi^{1/2}$ using $k_F=(3\pi^2n_e)^{1/3}$ where $n_e$ is the electron density. To control finite size effects we cutoff this potential at $r_{ij}=5\lambda$.  We use a step size of $\Delta t=3000$ fm/c.

\begin{figure}[htb]
\centering  
\includegraphics[width=0.47\textwidth]{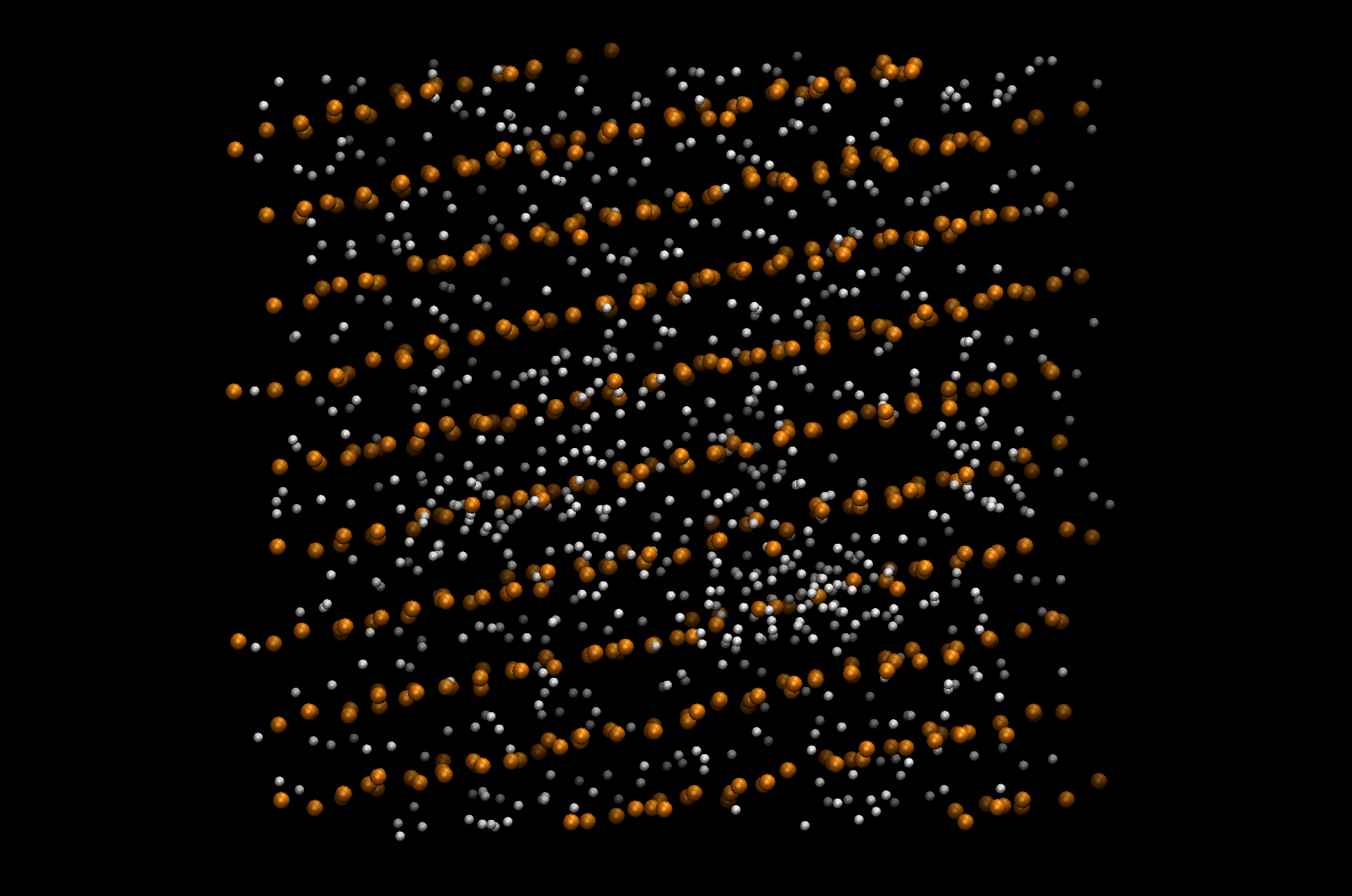}
\caption{\label{Fig1} Molecular dynamics simulation of a solid phase using 512 U ions (orange), 384 C (white) and 384 O ions (also white).}	
\end{figure}

We prepared an initial solid phase by starting a small system with 512 U, 384 C, and 384 O ions in a cubical simulation volume.  The initial positions were distributed randomly with a uniform probability.  The ion density was $7.34\times 10^{-10}$ fm$^{-3}$ and the temperature was 19 keV. The system was observed to freeze and the final configuration, after a simulation time of about $7.7\times 10^9$ fm/c, is shown in Fig. \ref{Fig1}. 

The U ions are approximately arranged on a crystal lattice as are some of the C and O ions.  However the remainder of the C and O ions appear to cluster in a nonuniform way.  We interpret this clustering as initial evidence for phase separation.  Some of the C/O appear to be starting to diffuse out of the crystal lattice.  This suggests that the equilibrium abundance of U in the solid phase is larger than our initial U abundance of 0.4.  We confirm this result below with a two phase simulation. 

\begin{figure}[htb]
\centering  
\includegraphics[width=0.47\textwidth]{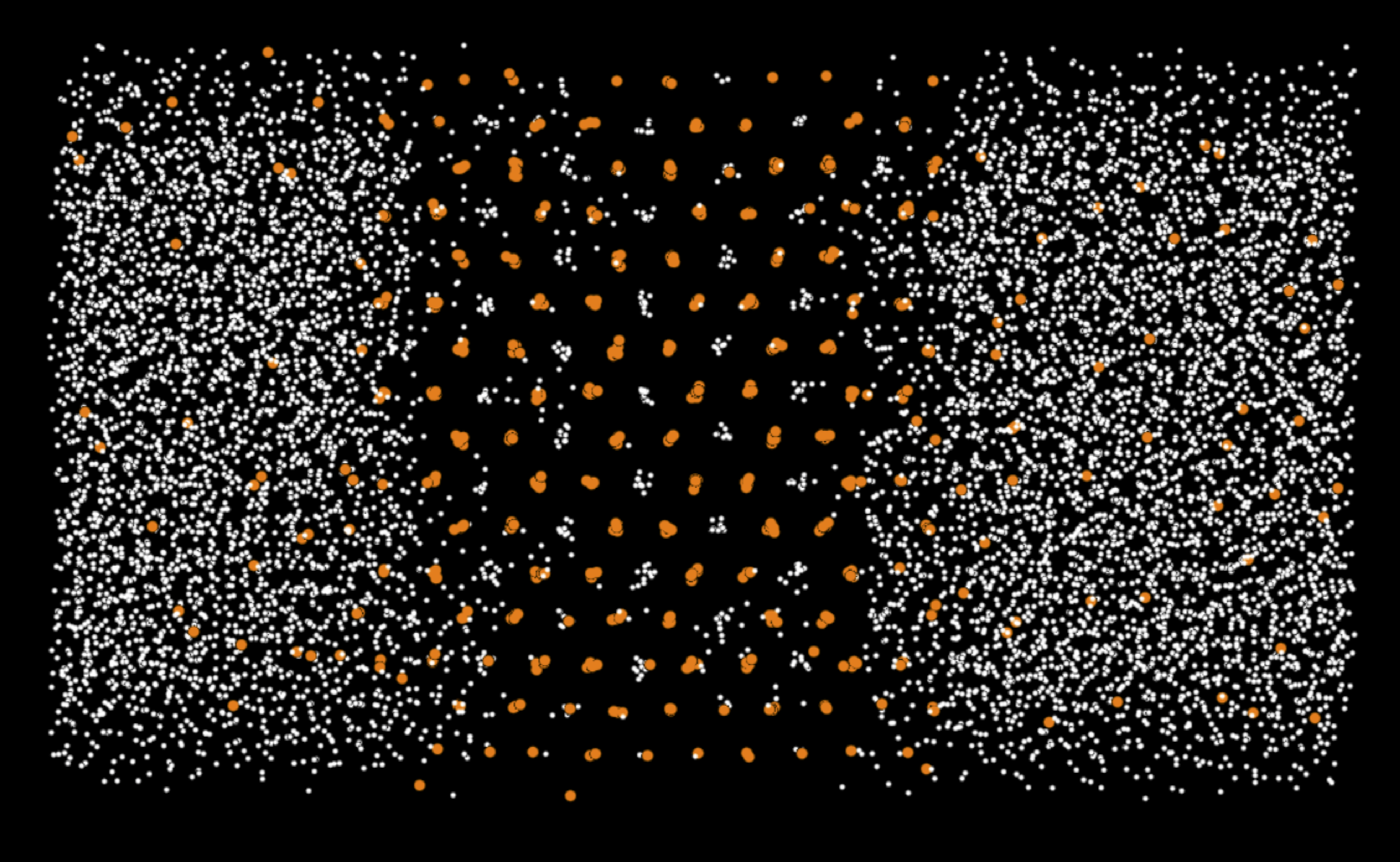}
\caption{\label{Fig1b} Molecular dynamics simulation of liquid / solid phase equilibrium using 8777 ions.  Uranium ions are orange while C and O ions are white.}	
\end{figure}

We next prepared a liquid phase with 3748 C and 3749 O ions in a cubical simulation volume of the same size and approximately the same electron density as the solid simulation. This simulation was equilibrated by running for a short time and then both simulations had there temperatures reduced to 5 keV.

The two systems were combined in a single rectangular shaped simulation volume that is twice as long in the z direction as in the x and y.  This combined system with a total of 8777 ions was evolved for a further $\approx 5\times 10^9$ fm/c at a temperature of 5 keV and than the somewhat high temperature of 7 keV to speed the diffusion of C and O.   The final configuration is shown in Fig.~\ref{Fig1b}.  The composition of this configuration is determined by dividing the simulation volume along the long axis into 20 equal subvolumes (slices) and determining the compsotion of each sub-volume by simply counting the number of ions of each type that are inside.  The results are shown in Fig.~\ref{Fig2}.

\begin{figure}[htb]
\centering  
\includegraphics[width=0.47\textwidth]{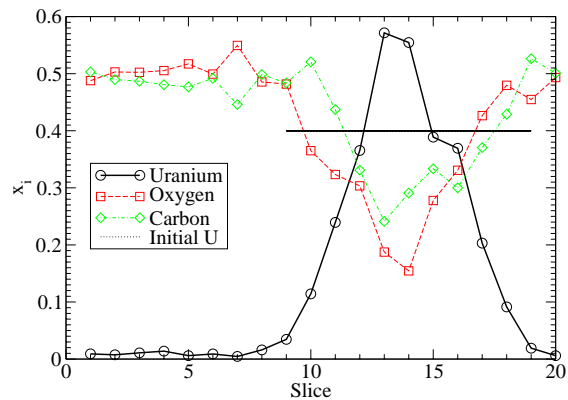}
\caption{\label{Fig2} Composition of the MD simulation described in the text versus sub-volume (slice) along the long axis of the rectangular simulation volume.  The U composition of the initial solid configuration was $x_U=0.40$ as indicated by the horizontal line.}	\end{figure}

The uranium concentration in the center of the solid phase is seen to increase from 0.4, the initial value of the original small solid simulation, to almost 0.6 as C and O diffused from the solid into the liquid.  This value 0.6 may represent something of a lower bound for the concentration of U in the solid.  Finite size effects and the limited running time of the simulation may not have allowed C and O to diffuse further possibly lowering the C and O concentration in the solid even further.  This will be checked with larger and longer MD simulations in future work.

\subsection{Alternate Lattices}

\begin{figure*}[htb]
\centering  
\includegraphics[width=1\textwidth]{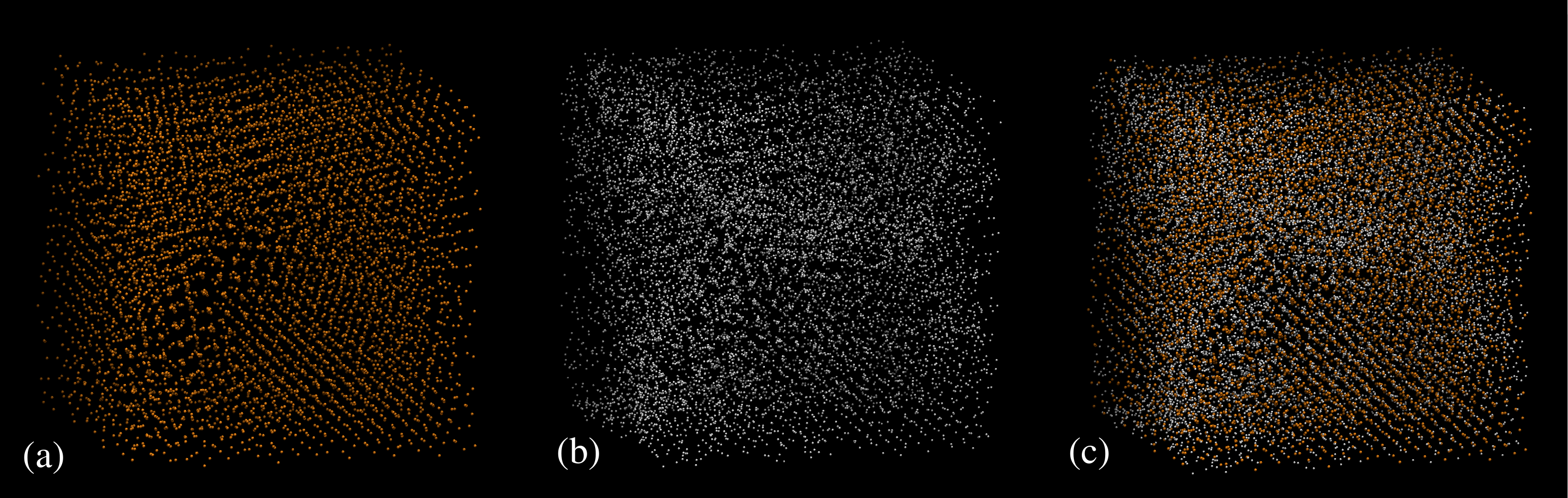}
\caption{\label{Fig3} Our mixture of 40\% C/O (white) and 60\% Pb/Th/U (orange) produces a complicated lattice when frozen. (a) With C/O omitted, the Pb/Th/U shows multiple crystal domains with a cubic lattice structure. Due to their high mass there is relatively little thermal noise on the lattice. (b) With Pb/Th/U omitted, it is more difficult to resolve the lattice with just C/O, but lattice planes of the domains are still apparent in some regions. The C/O clusters in filamentary walls along boundaries between the domains. (c) All nuclei. 
}	
\end{figure*}

The lattice structure of the solid may be complicated given the large charge ratios and is likely sensitive to the exact composition. We present here one example of a nontrivial lattice which formed spontaneously in an MD simulation. 

The mixture considered here contains 4142 C, 5760 O, 3588 Pb, 1238 Th, and 1656 U for a total of  16384 nuclei ($2^{14}$, for GPU threading), which is slightly enhanced in C and O relative to the solid described in Tab. \ref{Table1}. The mixture was initialized as a liquid with random initial positions with a temperature near $\Gamma_\mathrm{C} \approx 15$. At this $\Gamma$ the mixture was weakly supercooled and so it quickly crystallized. The crystal that formed had a CsCl type lattice, \ie\, a binary cubic lattice with a two-atom basis which alternates between actinides and C/O, and is shown in Fig. \ref{Fig3}. This is first example known to the authors of a lattice of this kind forming spontaneously in a simulation.

The number abundance of C/O is approximately 60\% while Pb/Th/U numbers about 40\%, which results a number of defects. The crystal appears to have multiple domains with different orientations, with a large number of C/O concentrated at the domain walls (arrows). This suggests the C/O abundance is too high and that a properly annealed crystal may expel light elements so that the ratio is very nearly 50/50 with actinides. In contrast, other lattice families such as orthorhombic or hexagonal may expel even more C/O and greatly enhance the actinide purity (see \cite{engstrom2016crystal} for examples). Therefore, if phase separation is sensitive to the exact lattice structure that nucleates then future work may consider lattice corrections to the solid free energy. 

Given the variation in charges on lattice sites there are likely screening effects, similar to what was reported in \cite{Caplan2020}. In short, we expect that there is likely some coordination so that the lowest Z may preferentially neighbor the highest Z so that unit cells may find an average charge to preserve long range order. We also resolve small clusters of C/O on some actinide lattice sites whose charge together acts like one actinide. Given the large timescale for actinide diffusion relative to C/O this configuration is difficult to anneal without a significant investment of computing time, but should be explored in detail in future work.


\begin{thebibliography}{35}%
\makeatletter
\providecommand \@ifxundefined [1]{%
 \@ifx{#1\undefined}
}%
\providecommand \@ifnum [1]{%
 \ifnum #1\expandafter \@firstoftwo
 \else \expandafter \@secondoftwo
 \fi
}%
\providecommand \@ifx [1]{%
 \ifx #1\expandafter \@firstoftwo
 \else \expandafter \@secondoftwo
 \fi
}%
\providecommand \natexlab [1]{#1}%
\providecommand \enquote  [1]{``#1''}%
\providecommand \bibnamefont  [1]{#1}%
\providecommand \bibfnamefont [1]{#1}%
\providecommand \citenamefont [1]{#1}%
\providecommand \href@noop [0]{\@secondoftwo}%
\providecommand \href [0]{\begingroup \@sanitize@url \@href}%
\providecommand \@href[1]{\@@startlink{#1}\@@href}%
\providecommand \@@href[1]{\endgroup#1\@@endlink}%
\providecommand \@sanitize@url [0]{\catcode `\\12\catcode `\$12\catcode
  `\&12\catcode `\#12\catcode `\^12\catcode `\_12\catcode `\%12\relax}%
\providecommand \@@startlink[1]{}%
\providecommand \@@endlink[0]{}%
\providecommand \url  [0]{\begingroup\@sanitize@url \@url }%
\providecommand \@url [1]{\endgroup\@href {#1}{\urlprefix }}%
\providecommand \urlprefix  [0]{URL }%
\providecommand \Eprint [0]{\href }%
\providecommand \doibase [0]{http://dx.doi.org/}%
\providecommand \selectlanguage [0]{\@gobble}%
\providecommand \bibinfo  [0]{\@secondoftwo}%
\providecommand \bibfield  [0]{\@secondoftwo}%
\providecommand \translation [1]{[#1]}%
\providecommand \BibitemOpen [0]{}%
\providecommand \bibitemStop [0]{}%
\providecommand \bibitemNoStop [0]{.\EOS\space}%
\providecommand \EOS [0]{\spacefactor3000\relax}%
\providecommand \BibitemShut  [1]{\csname bibitem#1\endcsname}%
\let\auto@bib@innerbib\@empty
\bibitem [{\citenamefont {{Gaia Collaboration}}(2018)}]{GaiaNoAuthors}%
  \BibitemOpen
  \bibfield  {author} {\bibinfo {author} {\bibnamefont {{Gaia
  Collaboration}}},\ }\href {\doibase 10.1051/0004-6361/201832843} {\bibfield
  {journal} {\bibinfo  {journal} {\aap}\ }\textbf {\bibinfo {volume} {616}},\
  \bibinfo {eid} {A10} (\bibinfo {year} {2018})},\ \Eprint
  {http://arxiv.org/abs/1804.09378} {arXiv:1804.09378 [astro-ph.SR]}
  \BibitemShut {NoStop}%
\bibitem [{\citenamefont {{van Horn}}(1968)}]{VanHorn1968}%
  \BibitemOpen
  \bibfield  {author} {\bibinfo {author} {\bibfnamefont {H.~M.}\ \bibnamefont
  {{van Horn}}},\ }\href {\doibase 10.1086/149432} {\bibfield  {journal}
  {\bibinfo  {journal} {\apj}\ }\textbf {\bibinfo {volume} {151}},\ \bibinfo
  {pages} {227} (\bibinfo {year} {1968})}\BibitemShut {NoStop}%
\bibitem [{\citenamefont {{Tremblay}}\ \emph {et~al.}(2019)\citenamefont
  {{Tremblay}}, \citenamefont {{Fontaine}}, \citenamefont {{Fusillo}},
  \citenamefont {{Dunlap}}, \citenamefont {{G{\"a}nsicke}}, \citenamefont
  {{Hollands}}, \citenamefont {{Hermes}}, \citenamefont {{Marsh}},
  \citenamefont {{Cukanovaite}},\ and\ \citenamefont
  {{Cunningham}}}]{Tremblay2019}%
  \BibitemOpen
  \bibfield  {author} {\bibinfo {author} {\bibfnamefont {P.-E.}\ \bibnamefont
  {{Tremblay}}}, \bibinfo {author} {\bibfnamefont {G.}~\bibnamefont
  {{Fontaine}}}, \bibinfo {author} {\bibfnamefont {N.~P.~G.}\ \bibnamefont
  {{Fusillo}}}, \bibinfo {author} {\bibfnamefont {B.~H.}\ \bibnamefont
  {{Dunlap}}}, \bibinfo {author} {\bibfnamefont {B.~T.}\ \bibnamefont
  {{G{\"a}nsicke}}}, \bibinfo {author} {\bibfnamefont {M.~A.}\ \bibnamefont
  {{Hollands}}}, \bibinfo {author} {\bibfnamefont {J.~J.}\ \bibnamefont
  {{Hermes}}}, \bibinfo {author} {\bibfnamefont {T.~R.}\ \bibnamefont
  {{Marsh}}}, \bibinfo {author} {\bibfnamefont {E.}~\bibnamefont
  {{Cukanovaite}}}, \ and\ \bibinfo {author} {\bibfnamefont {T.}~\bibnamefont
  {{Cunningham}}},\ }\href {\doibase 10.1038/s41586-018-0791-x} {\bibfield
  {journal} {\bibinfo  {journal} {\nat}\ }\textbf {\bibinfo {volume} {565}},\
  \bibinfo {pages} {202} (\bibinfo {year} {2019})},\ \Eprint
  {http://arxiv.org/abs/1908.00370} {arXiv:1908.00370 [astro-ph.SR]}
  \BibitemShut {NoStop}%
\bibitem [{\citenamefont {Bildsten}\ and\ \citenamefont
  {Hall}(2001)}]{Bildsten_2001}%
  \BibitemOpen
  \bibfield  {author} {\bibinfo {author} {\bibfnamefont {L.}~\bibnamefont
  {Bildsten}}\ and\ \bibinfo {author} {\bibfnamefont {D.~M.}\ \bibnamefont
  {Hall}},\ }\href {\doibase 10.1086/319169} {\bibfield  {journal} {\bibinfo
  {journal} {The Astrophysical Journal}\ }\textbf {\bibinfo {volume} {549}},\
  \bibinfo {pages} {L219} (\bibinfo {year} {2001})}\BibitemShut {NoStop}%
\bibitem [{\citenamefont {Hughto}\ \emph {et~al.}(2010)\citenamefont {Hughto},
  \citenamefont {Schneider}, \citenamefont {Horowitz},\ and\ \citenamefont
  {Berry}}]{PhysRevE.82.066401}%
  \BibitemOpen
  \bibfield  {author} {\bibinfo {author} {\bibfnamefont {J.}~\bibnamefont
  {Hughto}}, \bibinfo {author} {\bibfnamefont {A.~S.}\ \bibnamefont
  {Schneider}}, \bibinfo {author} {\bibfnamefont {C.~J.}\ \bibnamefont
  {Horowitz}}, \ and\ \bibinfo {author} {\bibfnamefont {D.~K.}\ \bibnamefont
  {Berry}},\ }\href {\doibase 10.1103/PhysRevE.82.066401} {\bibfield  {journal}
  {\bibinfo  {journal} {Phys. Rev. E}\ }\textbf {\bibinfo {volume} {82}},\
  \bibinfo {pages} {066401} (\bibinfo {year} {2010})}\BibitemShut {NoStop}%
\bibitem [{\citenamefont {{Blouin}}\ \emph {et~al.}(2020)\citenamefont
  {{Blouin}}, \citenamefont {{Daligault}}, \citenamefont {{Saumon}},
  \citenamefont {{B{\'e}dard}},\ and\ \citenamefont
  {{Brassard}}}]{2020arXiv200713669B}%
  \BibitemOpen
  \bibfield  {author} {\bibinfo {author} {\bibfnamefont {S.}~\bibnamefont
  {{Blouin}}}, \bibinfo {author} {\bibfnamefont {J.}~\bibnamefont
  {{Daligault}}}, \bibinfo {author} {\bibfnamefont {D.}~\bibnamefont
  {{Saumon}}}, \bibinfo {author} {\bibfnamefont {A.}~\bibnamefont
  {{B{\'e}dard}}}, \ and\ \bibinfo {author} {\bibfnamefont {P.}~\bibnamefont
  {{Brassard}}},\ }\href@noop {} {\bibfield  {journal} {\bibinfo  {journal}
  {arXiv e-prints}\ ,\ \bibinfo {eid} {arXiv:2007.13669}} (\bibinfo {year}
  {2020})},\ \Eprint {http://arxiv.org/abs/2007.13669} {arXiv:2007.13669
  [astro-ph.SR]} \BibitemShut {NoStop}%
\bibitem [{\citenamefont {{Camisassa}}\ \emph {et~al.}(2020)\citenamefont
  {{Camisassa}}, \citenamefont {{Althaus}}, \citenamefont {{Torres}},
  \citenamefont {{C{\'o}rsico}}, \citenamefont {{Cheng}},\ and\ \citenamefont
  {{Rebassa-Mansergas}}}]{Camisassa2020}%
  \BibitemOpen
  \bibfield  {author} {\bibinfo {author} {\bibfnamefont {M.~E.}\ \bibnamefont
  {{Camisassa}}}, \bibinfo {author} {\bibfnamefont {L.~G.}\ \bibnamefont
  {{Althaus}}}, \bibinfo {author} {\bibfnamefont {S.}~\bibnamefont {{Torres}}},
  \bibinfo {author} {\bibfnamefont {A.~H.}\ \bibnamefont {{C{\'o}rsico}}},
  \bibinfo {author} {\bibfnamefont {S.}~\bibnamefont {{Cheng}}}, \ and\
  \bibinfo {author} {\bibfnamefont {A.}~\bibnamefont {{Rebassa-Mansergas}}},\
  }\href@noop {} {\bibfield  {journal} {\bibinfo  {journal} {arXiv e-prints}\
  ,\ \bibinfo {eid} {arXiv:2008.03028}} (\bibinfo {year} {2020})},\ \Eprint
  {http://arxiv.org/abs/2008.03028} {arXiv:2008.03028 [astro-ph.SR]}
  \BibitemShut {NoStop}%
\bibitem [{\citenamefont {Cheng}\ \emph {et~al.}(2019)\citenamefont {Cheng},
  \citenamefont {Cummings},\ and\ \citenamefont {M{\'{e}}nard}}]{Cheng_2019}%
  \BibitemOpen
  \bibfield  {author} {\bibinfo {author} {\bibfnamefont {S.}~\bibnamefont
  {Cheng}}, \bibinfo {author} {\bibfnamefont {J.~D.}\ \bibnamefont {Cummings}},
  \ and\ \bibinfo {author} {\bibfnamefont {B.}~\bibnamefont {M{\'{e}}nard}},\
  }\href {\doibase 10.3847/1538-4357/ab4989} {\bibfield  {journal} {\bibinfo
  {journal} {The Astrophysical Journal}\ }\textbf {\bibinfo {volume} {886}},\
  \bibinfo {pages} {100} (\bibinfo {year} {2019})}\BibitemShut {NoStop}%
\bibitem [{\citenamefont {Bauer}\ \emph {et~al.}(2020)\citenamefont {Bauer},
  \citenamefont {Schwab}, \citenamefont {Bildsten},\ and\ \citenamefont
  {Cheng}}]{bauer2020multigigayear}%
  \BibitemOpen
  \bibfield  {author} {\bibinfo {author} {\bibfnamefont {E.~B.}\ \bibnamefont
  {Bauer}}, \bibinfo {author} {\bibfnamefont {J.}~\bibnamefont {Schwab}},
  \bibinfo {author} {\bibfnamefont {L.}~\bibnamefont {Bildsten}}, \ and\
  \bibinfo {author} {\bibfnamefont {S.}~\bibnamefont {Cheng}},\ }\href@noop {}
  {\enquote {\bibinfo {title} {Multi-gigayear white dwarf cooling delays from
  clustering-enhanced gravitational sedimentation},}\ } (\bibinfo {year}
  {2020}),\ \Eprint {http://arxiv.org/abs/2009.04025} {arXiv:2009.04025
  [astro-ph.SR]} \BibitemShut {NoStop}%
\bibitem [{\citenamefont {Caplan}\ \emph {et~al.}(2020)\citenamefont {Caplan},
  \citenamefont {Horowitz},\ and\ \citenamefont {Cumming}}]{Caplan_2020}%
  \BibitemOpen
  \bibfield  {author} {\bibinfo {author} {\bibfnamefont {M.~E.}\ \bibnamefont
  {Caplan}}, \bibinfo {author} {\bibfnamefont {C.~J.}\ \bibnamefont
  {Horowitz}}, \ and\ \bibinfo {author} {\bibfnamefont {A.}~\bibnamefont
  {Cumming}},\ }\href {\doibase 10.3847/2041-8213/abbda0} {\bibfield  {journal}
  {\bibinfo  {journal} {The Astrophysical Journal}\ }\textbf {\bibinfo {volume}
  {902}},\ \bibinfo {pages} {L44} (\bibinfo {year} {2020})}\BibitemShut
  {NoStop}%
\bibitem [{\citenamefont {Gauthier-Lafaye}\ \emph {et~al.}(1996)\citenamefont
  {Gauthier-Lafaye}, \citenamefont {Holliger},\ and\ \citenamefont
  {Blanc}}]{GAUTHIERLAFAYE19964831}%
  \BibitemOpen
  \bibfield  {author} {\bibinfo {author} {\bibfnamefont {F.}~\bibnamefont
  {Gauthier-Lafaye}}, \bibinfo {author} {\bibfnamefont {P.}~\bibnamefont
  {Holliger}}, \ and\ \bibinfo {author} {\bibfnamefont {P.-L.}\ \bibnamefont
  {Blanc}},\ }\href {\doibase https://doi.org/10.1016/S0016-7037(96)00245-1}
  {\bibfield  {journal} {\bibinfo  {journal} {Geochimica et Cosmochimica Acta}\
  }\textbf {\bibinfo {volume} {60}},\ \bibinfo {pages} {4831 } (\bibinfo {year}
  {1996})}\BibitemShut {NoStop}%
\bibitem [{\citenamefont {Kuroda}(1956)}]{doi:10.1063/1.1743058}%
  \BibitemOpen
  \bibfield  {author} {\bibinfo {author} {\bibfnamefont {P.~K.}\ \bibnamefont
  {Kuroda}},\ }\href {\doibase 10.1063/1.1743058} {\bibfield  {journal}
  {\bibinfo  {journal} {The Journal of Chemical Physics}\ }\textbf {\bibinfo
  {volume} {25}},\ \bibinfo {pages} {781} (\bibinfo {year} {1956})},\ \Eprint
  {http://arxiv.org/abs/https://doi.org/10.1063/1.1743058}
  {https://doi.org/10.1063/1.1743058} \BibitemShut {NoStop}%
\bibitem [{\citenamefont {Abbott}\ \emph {et~al.}(2019)\citenamefont {Abbott},
  \citenamefont {Allam}, \citenamefont {Andersen}, \citenamefont {Angus},
  \citenamefont {Asorey}, \citenamefont {Avelino}, \citenamefont {Avila},
  \citenamefont {Bassett}, \citenamefont {Bechtol}, \citenamefont {Bernstein},
  \citenamefont {Bertin}, \citenamefont {Brooks}, \citenamefont {Brout},
  \citenamefont {Brown}, \citenamefont {Burke}, \citenamefont {Calcino},
  \citenamefont {Rosell}, \citenamefont {Carollo}, \citenamefont {Kind},
  \citenamefont {Carretero}, \citenamefont {Casas}, \citenamefont {Castander},
  \citenamefont {Cawthon}, \citenamefont {Challis}, \citenamefont {Childress},
  \citenamefont {Clocchiatti}, \citenamefont {Cunha}, \citenamefont {D'Andrea},
  \citenamefont {da~Costa}, \citenamefont {Davis}, \citenamefont {Davis},
  \citenamefont {Vicente}, \citenamefont {DePoy}, \citenamefont {Desai},
  \citenamefont {Diehl}, \citenamefont {Doel}, \citenamefont {Drlica-Wagner},
  \citenamefont {Eifler}, \citenamefont {Evrard}, \citenamefont {Fernandez},
  \citenamefont {Filippenko}, \citenamefont {Finley}, \citenamefont {Flaugher},
  \citenamefont {Foley}, \citenamefont {Fosalba}, \citenamefont {Frieman},
  \citenamefont {Galbany}, \citenamefont {Garc{\'{\i}}a-Bellido}, \citenamefont
  {Gaztanaga}, \citenamefont {Giannantonio}, \citenamefont {Glazebrook},
  \citenamefont {Goldstein}, \citenamefont {Gonz{\'{a}}lez-Gait{\'{a}}n},
  \citenamefont {Gruen}, \citenamefont {Gruendl}, \citenamefont {Gschwend},
  \citenamefont {Gupta}, \citenamefont {Gutierrez}, \citenamefont {Hartley},
  \citenamefont {Hinton}, \citenamefont {Hollowood}, \citenamefont {Honscheid},
  \citenamefont {Hoormann}, \citenamefont {Hoyle}, \citenamefont {James},
  \citenamefont {Jeltema}, \citenamefont {Johnson}, \citenamefont {Johnson},
  \citenamefont {Kasai}, \citenamefont {Kent}, \citenamefont {Kessler},
  \citenamefont {Kim}, \citenamefont {Kirshner}, \citenamefont {Kovacs},
  \citenamefont {Krause}, \citenamefont {Kron}, \citenamefont {Kuehn},
  \citenamefont {Kuhlmann}, \citenamefont {Kuropatkin}, \citenamefont {Lahav},
  \citenamefont {Lasker}, \citenamefont {Lewis}, \citenamefont {Li},
  \citenamefont {Lidman}, \citenamefont {Lima}, \citenamefont {Lin},
  \citenamefont {Macaulay}, \citenamefont {Maia}, \citenamefont {Mandel},
  \citenamefont {March}, \citenamefont {Marriner}, \citenamefont {Marshall},
  \citenamefont {Martini}, \citenamefont {Menanteau}, \citenamefont {Miller},
  \citenamefont {Miquel}, \citenamefont {Miranda}, \citenamefont {Mohr},
  \citenamefont {Morganson}, \citenamefont {Muthukrishna}, \citenamefont
  {Möller}, \citenamefont {Neilsen}, \citenamefont {Nichol}, \citenamefont
  {Nord}, \citenamefont {Nugent}, \citenamefont {Ogando}, \citenamefont
  {Palmese}, \citenamefont {Pan}, \citenamefont {Plazas}, \citenamefont
  {Pursiainen}, \citenamefont {Romer}, \citenamefont {Roodman}, \citenamefont
  {Rozo}, \citenamefont {Rykoff}, \citenamefont {Sako}, \citenamefont
  {Sanchez}, \citenamefont {Scarpine}, \citenamefont {Schindler}, \citenamefont
  {Schubnell}, \citenamefont {Scolnic}, \citenamefont {Serrano}, \citenamefont
  {Sevilla-Noarbe}, \citenamefont {Sharp}, \citenamefont {Smith}, \citenamefont
  {Soares-Santos}, \citenamefont {Sobreira}, \citenamefont {Sommer},
  \citenamefont {Spinka}, \citenamefont {Suchyta}, \citenamefont {Sullivan},
  \citenamefont {Swann}, \citenamefont {Tarle}, \citenamefont {Thomas},
  \citenamefont {Thomas}, \citenamefont {Troxel}, \citenamefont {Tucker},
  \citenamefont {Uddin}, \citenamefont {Walker}, \citenamefont {Wester},
  \citenamefont {Wiseman}, \citenamefont {Wolf}, \citenamefont {Yanny},
  \citenamefont {Zhang},\ and\ \citenamefont {and}}]{Abbott_2019}%
  \BibitemOpen
  \bibfield  {author} {\bibinfo {author} {\bibfnamefont {T.~M.~C.}\
  \bibnamefont {Abbott}}, \bibinfo {author} {\bibfnamefont {S.}~\bibnamefont
  {Allam}}, \bibinfo {author} {\bibfnamefont {P.}~\bibnamefont {Andersen}},
  \bibinfo {author} {\bibfnamefont {C.}~\bibnamefont {Angus}}, \bibinfo
  {author} {\bibfnamefont {J.}~\bibnamefont {Asorey}}, \bibinfo {author}
  {\bibfnamefont {A.}~\bibnamefont {Avelino}}, \bibinfo {author} {\bibfnamefont
  {S.}~\bibnamefont {Avila}}, \bibinfo {author} {\bibfnamefont {B.~A.}\
  \bibnamefont {Bassett}}, \bibinfo {author} {\bibfnamefont {K.}~\bibnamefont
  {Bechtol}}, \bibinfo {author} {\bibfnamefont {G.~M.}\ \bibnamefont
  {Bernstein}}, \bibinfo {author} {\bibfnamefont {E.}~\bibnamefont {Bertin}},
  \bibinfo {author} {\bibfnamefont {D.}~\bibnamefont {Brooks}}, \bibinfo
  {author} {\bibfnamefont {D.}~\bibnamefont {Brout}}, \bibinfo {author}
  {\bibfnamefont {P.}~\bibnamefont {Brown}}, \bibinfo {author} {\bibfnamefont
  {D.~L.}\ \bibnamefont {Burke}}, \bibinfo {author} {\bibfnamefont
  {J.}~\bibnamefont {Calcino}}, \bibinfo {author} {\bibfnamefont {A.~C.}\
  \bibnamefont {Rosell}}, \bibinfo {author} {\bibfnamefont {D.}~\bibnamefont
  {Carollo}}, \bibinfo {author} {\bibfnamefont {M.~C.}\ \bibnamefont {Kind}},
  \bibinfo {author} {\bibfnamefont {J.}~\bibnamefont {Carretero}}, \bibinfo
  {author} {\bibfnamefont {R.}~\bibnamefont {Casas}}, \bibinfo {author}
  {\bibfnamefont {F.~J.}\ \bibnamefont {Castander}}, \bibinfo {author}
  {\bibfnamefont {R.}~\bibnamefont {Cawthon}}, \bibinfo {author} {\bibfnamefont
  {P.}~\bibnamefont {Challis}}, \bibinfo {author} {\bibfnamefont
  {M.}~\bibnamefont {Childress}}, \bibinfo {author} {\bibfnamefont
  {A.}~\bibnamefont {Clocchiatti}}, \bibinfo {author} {\bibfnamefont {C.~E.}\
  \bibnamefont {Cunha}}, \bibinfo {author} {\bibfnamefont {C.~B.}\ \bibnamefont
  {D'Andrea}}, \bibinfo {author} {\bibfnamefont {L.~N.}\ \bibnamefont
  {da~Costa}}, \bibinfo {author} {\bibfnamefont {C.}~\bibnamefont {Davis}},
  \bibinfo {author} {\bibfnamefont {T.~M.}\ \bibnamefont {Davis}}, \bibinfo
  {author} {\bibfnamefont {J.~D.}\ \bibnamefont {Vicente}}, \bibinfo {author}
  {\bibfnamefont {D.~L.}\ \bibnamefont {DePoy}}, \bibinfo {author}
  {\bibfnamefont {S.}~\bibnamefont {Desai}}, \bibinfo {author} {\bibfnamefont
  {H.~T.}\ \bibnamefont {Diehl}}, \bibinfo {author} {\bibfnamefont
  {P.}~\bibnamefont {Doel}}, \bibinfo {author} {\bibfnamefont {A.}~\bibnamefont
  {Drlica-Wagner}}, \bibinfo {author} {\bibfnamefont {T.~F.}\ \bibnamefont
  {Eifler}}, \bibinfo {author} {\bibfnamefont {A.~E.}\ \bibnamefont {Evrard}},
  \bibinfo {author} {\bibfnamefont {E.}~\bibnamefont {Fernandez}}, \bibinfo
  {author} {\bibfnamefont {A.~V.}\ \bibnamefont {Filippenko}}, \bibinfo
  {author} {\bibfnamefont {D.~A.}\ \bibnamefont {Finley}}, \bibinfo {author}
  {\bibfnamefont {B.}~\bibnamefont {Flaugher}}, \bibinfo {author}
  {\bibfnamefont {R.~J.}\ \bibnamefont {Foley}}, \bibinfo {author}
  {\bibfnamefont {P.}~\bibnamefont {Fosalba}}, \bibinfo {author} {\bibfnamefont
  {J.}~\bibnamefont {Frieman}}, \bibinfo {author} {\bibfnamefont
  {L.}~\bibnamefont {Galbany}}, \bibinfo {author} {\bibfnamefont
  {J.}~\bibnamefont {Garc{\'{\i}}a-Bellido}}, \bibinfo {author} {\bibfnamefont
  {E.}~\bibnamefont {Gaztanaga}}, \bibinfo {author} {\bibfnamefont
  {T.}~\bibnamefont {Giannantonio}}, \bibinfo {author} {\bibfnamefont
  {K.}~\bibnamefont {Glazebrook}}, \bibinfo {author} {\bibfnamefont {D.~A.}\
  \bibnamefont {Goldstein}}, \bibinfo {author} {\bibfnamefont {S.}~\bibnamefont
  {Gonz{\'{a}}lez-Gait{\'{a}}n}}, \bibinfo {author} {\bibfnamefont
  {D.}~\bibnamefont {Gruen}}, \bibinfo {author} {\bibfnamefont {R.~A.}\
  \bibnamefont {Gruendl}}, \bibinfo {author} {\bibfnamefont {J.}~\bibnamefont
  {Gschwend}}, \bibinfo {author} {\bibfnamefont {R.~R.}\ \bibnamefont {Gupta}},
  \bibinfo {author} {\bibfnamefont {G.}~\bibnamefont {Gutierrez}}, \bibinfo
  {author} {\bibfnamefont {W.~G.}\ \bibnamefont {Hartley}}, \bibinfo {author}
  {\bibfnamefont {S.~R.}\ \bibnamefont {Hinton}}, \bibinfo {author}
  {\bibfnamefont {D.~L.}\ \bibnamefont {Hollowood}}, \bibinfo {author}
  {\bibfnamefont {K.}~\bibnamefont {Honscheid}}, \bibinfo {author}
  {\bibfnamefont {J.~K.}\ \bibnamefont {Hoormann}}, \bibinfo {author}
  {\bibfnamefont {B.}~\bibnamefont {Hoyle}}, \bibinfo {author} {\bibfnamefont
  {D.~J.}\ \bibnamefont {James}}, \bibinfo {author} {\bibfnamefont
  {T.}~\bibnamefont {Jeltema}}, \bibinfo {author} {\bibfnamefont {M.~W.~G.}\
  \bibnamefont {Johnson}}, \bibinfo {author} {\bibfnamefont {M.~D.}\
  \bibnamefont {Johnson}}, \bibinfo {author} {\bibfnamefont {E.}~\bibnamefont
  {Kasai}}, \bibinfo {author} {\bibfnamefont {S.}~\bibnamefont {Kent}},
  \bibinfo {author} {\bibfnamefont {R.}~\bibnamefont {Kessler}}, \bibinfo
  {author} {\bibfnamefont {A.~G.}\ \bibnamefont {Kim}}, \bibinfo {author}
  {\bibfnamefont {R.~P.}\ \bibnamefont {Kirshner}}, \bibinfo {author}
  {\bibfnamefont {E.}~\bibnamefont {Kovacs}}, \bibinfo {author} {\bibfnamefont
  {E.}~\bibnamefont {Krause}}, \bibinfo {author} {\bibfnamefont
  {R.}~\bibnamefont {Kron}}, \bibinfo {author} {\bibfnamefont {K.}~\bibnamefont
  {Kuehn}}, \bibinfo {author} {\bibfnamefont {S.}~\bibnamefont {Kuhlmann}},
  \bibinfo {author} {\bibfnamefont {N.}~\bibnamefont {Kuropatkin}}, \bibinfo
  {author} {\bibfnamefont {O.}~\bibnamefont {Lahav}}, \bibinfo {author}
  {\bibfnamefont {J.}~\bibnamefont {Lasker}}, \bibinfo {author} {\bibfnamefont
  {G.~F.}\ \bibnamefont {Lewis}}, \bibinfo {author} {\bibfnamefont {T.~S.}\
  \bibnamefont {Li}}, \bibinfo {author} {\bibfnamefont {C.}~\bibnamefont
  {Lidman}}, \bibinfo {author} {\bibfnamefont {M.}~\bibnamefont {Lima}},
  \bibinfo {author} {\bibfnamefont {H.}~\bibnamefont {Lin}}, \bibinfo {author}
  {\bibfnamefont {E.}~\bibnamefont {Macaulay}}, \bibinfo {author}
  {\bibfnamefont {M.~A.~G.}\ \bibnamefont {Maia}}, \bibinfo {author}
  {\bibfnamefont {K.~S.}\ \bibnamefont {Mandel}}, \bibinfo {author}
  {\bibfnamefont {M.}~\bibnamefont {March}}, \bibinfo {author} {\bibfnamefont
  {J.}~\bibnamefont {Marriner}}, \bibinfo {author} {\bibfnamefont {J.~L.}\
  \bibnamefont {Marshall}}, \bibinfo {author} {\bibfnamefont {P.}~\bibnamefont
  {Martini}}, \bibinfo {author} {\bibfnamefont {F.}~\bibnamefont {Menanteau}},
  \bibinfo {author} {\bibfnamefont {C.~J.}\ \bibnamefont {Miller}}, \bibinfo
  {author} {\bibfnamefont {R.}~\bibnamefont {Miquel}}, \bibinfo {author}
  {\bibfnamefont {V.}~\bibnamefont {Miranda}}, \bibinfo {author} {\bibfnamefont
  {J.~J.}\ \bibnamefont {Mohr}}, \bibinfo {author} {\bibfnamefont
  {E.}~\bibnamefont {Morganson}}, \bibinfo {author} {\bibfnamefont
  {D.}~\bibnamefont {Muthukrishna}}, \bibinfo {author} {\bibfnamefont
  {A.}~\bibnamefont {Möller}}, \bibinfo {author} {\bibfnamefont
  {E.}~\bibnamefont {Neilsen}}, \bibinfo {author} {\bibfnamefont {R.~C.}\
  \bibnamefont {Nichol}}, \bibinfo {author} {\bibfnamefont {B.}~\bibnamefont
  {Nord}}, \bibinfo {author} {\bibfnamefont {P.}~\bibnamefont {Nugent}},
  \bibinfo {author} {\bibfnamefont {R.~L.~C.}\ \bibnamefont {Ogando}}, \bibinfo
  {author} {\bibfnamefont {A.}~\bibnamefont {Palmese}}, \bibinfo {author}
  {\bibfnamefont {Y.-C.}\ \bibnamefont {Pan}}, \bibinfo {author} {\bibfnamefont
  {A.~A.}\ \bibnamefont {Plazas}}, \bibinfo {author} {\bibfnamefont
  {M.}~\bibnamefont {Pursiainen}}, \bibinfo {author} {\bibfnamefont {A.~K.}\
  \bibnamefont {Romer}}, \bibinfo {author} {\bibfnamefont {A.}~\bibnamefont
  {Roodman}}, \bibinfo {author} {\bibfnamefont {E.}~\bibnamefont {Rozo}},
  \bibinfo {author} {\bibfnamefont {E.~S.}\ \bibnamefont {Rykoff}}, \bibinfo
  {author} {\bibfnamefont {M.}~\bibnamefont {Sako}}, \bibinfo {author}
  {\bibfnamefont {E.}~\bibnamefont {Sanchez}}, \bibinfo {author} {\bibfnamefont
  {V.}~\bibnamefont {Scarpine}}, \bibinfo {author} {\bibfnamefont
  {R.}~\bibnamefont {Schindler}}, \bibinfo {author} {\bibfnamefont
  {M.}~\bibnamefont {Schubnell}}, \bibinfo {author} {\bibfnamefont
  {D.}~\bibnamefont {Scolnic}}, \bibinfo {author} {\bibfnamefont
  {S.}~\bibnamefont {Serrano}}, \bibinfo {author} {\bibfnamefont
  {I.}~\bibnamefont {Sevilla-Noarbe}}, \bibinfo {author} {\bibfnamefont
  {R.}~\bibnamefont {Sharp}}, \bibinfo {author} {\bibfnamefont
  {M.}~\bibnamefont {Smith}}, \bibinfo {author} {\bibfnamefont
  {M.}~\bibnamefont {Soares-Santos}}, \bibinfo {author} {\bibfnamefont
  {F.}~\bibnamefont {Sobreira}}, \bibinfo {author} {\bibfnamefont {N.~E.}\
  \bibnamefont {Sommer}}, \bibinfo {author} {\bibfnamefont {H.}~\bibnamefont
  {Spinka}}, \bibinfo {author} {\bibfnamefont {E.}~\bibnamefont {Suchyta}},
  \bibinfo {author} {\bibfnamefont {M.}~\bibnamefont {Sullivan}}, \bibinfo
  {author} {\bibfnamefont {E.}~\bibnamefont {Swann}}, \bibinfo {author}
  {\bibfnamefont {G.}~\bibnamefont {Tarle}}, \bibinfo {author} {\bibfnamefont
  {D.}~\bibnamefont {Thomas}}, \bibinfo {author} {\bibfnamefont {R.~C.}\
  \bibnamefont {Thomas}}, \bibinfo {author} {\bibfnamefont {M.~A.}\
  \bibnamefont {Troxel}}, \bibinfo {author} {\bibfnamefont {B.~E.}\
  \bibnamefont {Tucker}}, \bibinfo {author} {\bibfnamefont {S.~A.}\
  \bibnamefont {Uddin}}, \bibinfo {author} {\bibfnamefont {A.~R.}\ \bibnamefont
  {Walker}}, \bibinfo {author} {\bibfnamefont {W.}~\bibnamefont {Wester}},
  \bibinfo {author} {\bibfnamefont {P.}~\bibnamefont {Wiseman}}, \bibinfo
  {author} {\bibfnamefont {R.~C.}\ \bibnamefont {Wolf}}, \bibinfo {author}
  {\bibfnamefont {B.}~\bibnamefont {Yanny}}, \bibinfo {author} {\bibfnamefont
  {B.}~\bibnamefont {Zhang}}, \ and\ \bibinfo {author} {\bibfnamefont {Y.~Z.}\
  \bibnamefont {and}},\ }\href {\doibase 10.3847/2041-8213/ab04fa} {\bibfield
  {journal} {\bibinfo  {journal} {The Astrophysical Journal}\ }\textbf
  {\bibinfo {volume} {872}},\ \bibinfo {pages} {L30} (\bibinfo {year}
  {2019})}\BibitemShut {NoStop}%
\bibitem [{\citenamefont {Howell}(2011)}]{SN_cosmology}%
  \BibitemOpen
  \bibfield  {author} {\bibinfo {author} {\bibfnamefont {D.~A.}\ \bibnamefont
  {Howell}},\ }\href {\doibase 10.1038/ncomms1344} {\bibfield  {journal}
  {\bibinfo  {journal} {Nature Communications}\ }\textbf {\bibinfo {volume}
  {2}},\ \bibinfo {pages} {350} (\bibinfo {year} {2011})}\BibitemShut {NoStop}%
\bibitem [{\citenamefont {Sullivan}(2010)}]{Sullivan2010}%
  \BibitemOpen
  \bibfield  {author} {\bibinfo {author} {\bibfnamefont {M.}~\bibnamefont
  {Sullivan}},\ }\enquote {\bibinfo {title} {Type ia supernovae and
  cosmology},}\ in\ \href {\doibase 10.1007/978-3-642-10598-2_2} {\emph
  {\bibinfo {booktitle} {Lectures on Cosmology: Accelerated Expansion of the
  Universe}}},\ \bibinfo {editor} {edited by\ \bibinfo {editor} {\bibfnamefont
  {G.}~\bibnamefont {Wolschin}}}\ (\bibinfo  {publisher} {Springer Berlin
  Heidelberg},\ \bibinfo {address} {Berlin, Heidelberg},\ \bibinfo {year}
  {2010})\ pp.\ \bibinfo {pages} {59--97}\BibitemShut {NoStop}%
\bibitem [{\citenamefont {Perlmutter}\ \emph {et~al.}(2011)\citenamefont
  {Perlmutter}, \citenamefont {Schmidt},\ and\ \citenamefont {Riess}}]{Nobel}%
  \BibitemOpen
  \bibfield  {author} {\bibinfo {author} {\bibfnamefont {S.}~\bibnamefont
  {Perlmutter}}, \bibinfo {author} {\bibfnamefont {B.~P.}\ \bibnamefont
  {Schmidt}}, \ and\ \bibinfo {author} {\bibfnamefont {A.~G.}\ \bibnamefont
  {Riess}},\ }\href@noop {} {\enquote {\bibinfo {title} {The nobel prize in
  physics for the discovery of the accelerating expansion of the universe
  through observations of distant supernovae.}}\ } (\bibinfo {year}
  {2011})\BibitemShut {NoStop}%
\bibitem [{\citenamefont {{Wang}}\ and\ \citenamefont
  {{Han}}(2012)}]{2012NewAR..56..122W}%
  \BibitemOpen
  \bibfield  {author} {\bibinfo {author} {\bibfnamefont {B.}~\bibnamefont
  {{Wang}}}\ and\ \bibinfo {author} {\bibfnamefont {Z.}~\bibnamefont {{Han}}},\
  }\href {\doibase 10.1016/j.newar.2012.04.001} {\bibfield  {journal} {\bibinfo
   {journal} {New Astronomy Reviews}\ }\textbf {\bibinfo {volume} {56}},\
  \bibinfo {pages} {122} (\bibinfo {year} {2012})},\ \Eprint
  {http://arxiv.org/abs/1204.1155} {arXiv:1204.1155 [astro-ph.SR]} \BibitemShut
  {NoStop}%
\bibitem [{\citenamefont {Hillebrandt}\ \emph {et~al.}(2013)\citenamefont
  {Hillebrandt}, \citenamefont {Kromer}, \citenamefont {Röpke},\ and\
  \citenamefont {Ruiter}}]{hillebrandt2013understanding}%
  \BibitemOpen
  \bibfield  {author} {\bibinfo {author} {\bibfnamefont {W.}~\bibnamefont
  {Hillebrandt}}, \bibinfo {author} {\bibfnamefont {M.}~\bibnamefont {Kromer}},
  \bibinfo {author} {\bibfnamefont {F.~K.}\ \bibnamefont {Röpke}}, \ and\
  \bibinfo {author} {\bibfnamefont {A.~J.}\ \bibnamefont {Ruiter}},\
  }\href@noop {} {\enquote {\bibinfo {title} {Towards an understanding of type
  ia supernovae from a synthesis of theory and observations},}\ } (\bibinfo
  {year} {2013}),\ \Eprint {http://arxiv.org/abs/1302.6420} {arXiv:1302.6420
  [astro-ph.CO]} \BibitemShut {NoStop}%
\bibitem [{\citenamefont {Ruiz-Lapuente}(2014)}]{RUIZLAPUENTE201415}%
  \BibitemOpen
  \bibfield  {author} {\bibinfo {author} {\bibfnamefont {P.}~\bibnamefont
  {Ruiz-Lapuente}},\ }\href {\doibase
  https://doi.org/10.1016/j.newar.2014.08.002} {\bibfield  {journal} {\bibinfo
  {journal} {New Astronomy Reviews}\ }\textbf {\bibinfo {volume} {62-63}},\
  \bibinfo {pages} {15 } (\bibinfo {year} {2014})}\BibitemShut {NoStop}%
\bibitem [{\citenamefont {Mckinven}\ \emph {et~al.}(2016)\citenamefont
  {Mckinven}, \citenamefont {Cumming}, \citenamefont {Medin},\ and\
  \citenamefont {Schatz}}]{Mckinven_2016}%
  \BibitemOpen
  \bibfield  {author} {\bibinfo {author} {\bibfnamefont {R.}~\bibnamefont
  {Mckinven}}, \bibinfo {author} {\bibfnamefont {A.}~\bibnamefont {Cumming}},
  \bibinfo {author} {\bibfnamefont {Z.}~\bibnamefont {Medin}}, \ and\ \bibinfo
  {author} {\bibfnamefont {H.}~\bibnamefont {Schatz}},\ }\href {\doibase
  10.3847/0004-637x/823/2/117} {\bibfield  {journal} {\bibinfo  {journal} {The
  Astrophysical Journal}\ }\textbf {\bibinfo {volume} {823}},\ \bibinfo {pages}
  {117} (\bibinfo {year} {2016})}\BibitemShut {NoStop}%
\bibitem [{\citenamefont {Horowitz}\ \emph {et~al.}(2010)\citenamefont
  {Horowitz}, \citenamefont {Schneider},\ and\ \citenamefont
  {Berry}}]{PhysRevLett.104.231101}%
  \BibitemOpen
  \bibfield  {author} {\bibinfo {author} {\bibfnamefont {C.~J.}\ \bibnamefont
  {Horowitz}}, \bibinfo {author} {\bibfnamefont {A.~S.}\ \bibnamefont
  {Schneider}}, \ and\ \bibinfo {author} {\bibfnamefont {D.~K.}\ \bibnamefont
  {Berry}},\ }\href {\doibase 10.1103/PhysRevLett.104.231101} {\bibfield
  {journal} {\bibinfo  {journal} {Phys. Rev. Lett.}\ }\textbf {\bibinfo
  {volume} {104}},\ \bibinfo {pages} {231101} (\bibinfo {year}
  {2010})}\BibitemShut {NoStop}%
\bibitem [{\citenamefont {Medin}\ and\ \citenamefont
  {Cumming}(2010)}]{PhysRevE.81.036107}%
  \BibitemOpen
  \bibfield  {author} {\bibinfo {author} {\bibfnamefont {Z.}~\bibnamefont
  {Medin}}\ and\ \bibinfo {author} {\bibfnamefont {A.}~\bibnamefont
  {Cumming}},\ }\href {\doibase 10.1103/PhysRevE.81.036107} {\bibfield
  {journal} {\bibinfo  {journal} {Phys. Rev. E}\ }\textbf {\bibinfo {volume}
  {81}},\ \bibinfo {pages} {036107} (\bibinfo {year} {2010})}\BibitemShut
  {NoStop}%
\bibitem [{\citenamefont {Holmbeck}\ \emph {et~al.}(2018)\citenamefont
  {Holmbeck}, \citenamefont {Beers}, \citenamefont {Roederer}, \citenamefont
  {Placco}, \citenamefont {Hansen}, \citenamefont {Sakari}, \citenamefont
  {Sneden}, \citenamefont {Liu}, \citenamefont {Lee}, \citenamefont {Cowan},\
  and\ \citenamefont {Frebel}}]{Holmbeck_2018}%
  \BibitemOpen
  \bibfield  {author} {\bibinfo {author} {\bibfnamefont {E.~M.}\ \bibnamefont
  {Holmbeck}}, \bibinfo {author} {\bibfnamefont {T.~C.}\ \bibnamefont {Beers}},
  \bibinfo {author} {\bibfnamefont {I.~U.}\ \bibnamefont {Roederer}}, \bibinfo
  {author} {\bibfnamefont {V.~M.}\ \bibnamefont {Placco}}, \bibinfo {author}
  {\bibfnamefont {T.~T.}\ \bibnamefont {Hansen}}, \bibinfo {author}
  {\bibfnamefont {C.~M.}\ \bibnamefont {Sakari}}, \bibinfo {author}
  {\bibfnamefont {C.}~\bibnamefont {Sneden}}, \bibinfo {author} {\bibfnamefont
  {C.}~\bibnamefont {Liu}}, \bibinfo {author} {\bibfnamefont {Y.~S.}\
  \bibnamefont {Lee}}, \bibinfo {author} {\bibfnamefont {J.~J.}\ \bibnamefont
  {Cowan}}, \ and\ \bibinfo {author} {\bibfnamefont {A.}~\bibnamefont
  {Frebel}},\ }\href {\doibase 10.3847/2041-8213/aac722} {\bibfield  {journal}
  {\bibinfo  {journal} {The Astrophysical Journal}\ }\textbf {\bibinfo {volume}
  {859}},\ \bibinfo {pages} {L24} (\bibinfo {year} {2018})}\BibitemShut
  {NoStop}%
\bibitem [{\citenamefont {Lodders}(2019)}]{lodders2019solar}%
  \BibitemOpen
  \bibfield  {author} {\bibinfo {author} {\bibfnamefont {K.}~\bibnamefont
  {Lodders}},\ }\href@noop {} {\enquote {\bibinfo {title} {Solar elemental
  abundances},}\ } (\bibinfo {year} {2019}),\ \Eprint
  {http://arxiv.org/abs/1912.00844} {arXiv:1912.00844 [astro-ph.SR]}
  \BibitemShut {NoStop}%
\bibitem [{sup()}]{supplemental}%
  \BibitemOpen
  \href@noop {} {\bibinfo  {journal} {On line supplemental material for this
  article, see Appendix}\ }\BibitemShut {NoStop}%
\bibitem [{NND()}]{NNDC}%
  \BibitemOpen
\bibfield  {journal} {  }\href@noop {} {\bibinfo  {journal} {National Nuclear
  Data Center at http://www.nndc.bnl.gov}\ }\BibitemShut {NoStop}%
\bibitem [{\citenamefont {Lamarsh}(1977)}]{Lamarsh}%
  \BibitemOpen
\bibfield  {journal} {  }\bibfield  {author} {\bibinfo {author} {\bibfnamefont
  {J.~R.}\ \bibnamefont {Lamarsh}},\ }\href@noop {} {\emph {\bibinfo {title}
  {Introduction to Nuclear Engineering}}}\ (\bibinfo  {publisher} {Addison
  Wesley},\ \bibinfo {year} {1977})\BibitemShut {NoStop}%
\bibitem [{\citenamefont {Cummings}\ \emph {et~al.}(2018)\citenamefont
  {Cummings}, \citenamefont {Kalirai}, \citenamefont {Tremblay}, \citenamefont
  {Ramirez-Ruiz},\ and\ \citenamefont {Choi}}]{Cummings_2018}%
  \BibitemOpen
  \bibfield  {author} {\bibinfo {author} {\bibfnamefont {J.~D.}\ \bibnamefont
  {Cummings}}, \bibinfo {author} {\bibfnamefont {J.~S.}\ \bibnamefont
  {Kalirai}}, \bibinfo {author} {\bibfnamefont {P.-E.}\ \bibnamefont
  {Tremblay}}, \bibinfo {author} {\bibfnamefont {E.}~\bibnamefont
  {Ramirez-Ruiz}}, \ and\ \bibinfo {author} {\bibfnamefont {J.}~\bibnamefont
  {Choi}},\ }\href {\doibase 10.3847/1538-4357/aadfd6} {\bibfield  {journal}
  {\bibinfo  {journal} {The Astrophysical Journal}\ }\textbf {\bibinfo {volume}
  {866}},\ \bibinfo {pages} {21} (\bibinfo {year} {2018})}\BibitemShut
  {NoStop}%
\bibitem [{\citenamefont {Cooper}\ and\ \citenamefont
  {Bildsten}(2008)}]{PhysRevE.77.056405}%
  \BibitemOpen
  \bibfield  {author} {\bibinfo {author} {\bibfnamefont {R.~L.}\ \bibnamefont
  {Cooper}}\ and\ \bibinfo {author} {\bibfnamefont {L.}~\bibnamefont
  {Bildsten}},\ }\href {\doibase 10.1103/PhysRevE.77.056405} {\bibfield
  {journal} {\bibinfo  {journal} {Phys. Rev. E}\ }\textbf {\bibinfo {volume}
  {77}},\ \bibinfo {pages} {056405} (\bibinfo {year} {2008})}\BibitemShut
  {NoStop}%
\bibitem [{\citenamefont {Shternin}\ and\ \citenamefont
  {Yakovlev}(2006)}]{PhysRevD.74.043004}%
  \BibitemOpen
  \bibfield  {author} {\bibinfo {author} {\bibfnamefont {P.~S.}\ \bibnamefont
  {Shternin}}\ and\ \bibinfo {author} {\bibfnamefont {D.~G.}\ \bibnamefont
  {Yakovlev}},\ }\href {\doibase 10.1103/PhysRevD.74.043004} {\bibfield
  {journal} {\bibinfo  {journal} {Phys. Rev. D}\ }\textbf {\bibinfo {volume}
  {74}},\ \bibinfo {pages} {043004} (\bibinfo {year} {2006})}\BibitemShut
  {NoStop}%
\bibitem [{\citenamefont {{Timmes}}\ and\ \citenamefont
  {{Woosley}}(1992)}]{1992ApJ...396..649T}%
  \BibitemOpen
  \bibfield  {author} {\bibinfo {author} {\bibfnamefont {F.~X.}\ \bibnamefont
  {{Timmes}}}\ and\ \bibinfo {author} {\bibfnamefont {S.~E.}\ \bibnamefont
  {{Woosley}}},\ }\href {\doibase 10.1086/171746} {\bibfield  {journal}
  {\bibinfo  {journal} {\apj}\ }\textbf {\bibinfo {volume} {396}},\ \bibinfo
  {pages} {649} (\bibinfo {year} {1992})}\BibitemShut {NoStop}%
\bibitem [{\citenamefont {Poludnenko}\ \emph {et~al.}(2019)\citenamefont
  {Poludnenko}, \citenamefont {Chambers}, \citenamefont {Ahmed}, \citenamefont
  {Gamezo},\ and\ \citenamefont {Taylor}}]{Poludnenkoeaau7365}%
  \BibitemOpen
  \bibfield  {author} {\bibinfo {author} {\bibfnamefont {A.~Y.}\ \bibnamefont
  {Poludnenko}}, \bibinfo {author} {\bibfnamefont {J.}~\bibnamefont
  {Chambers}}, \bibinfo {author} {\bibfnamefont {K.}~\bibnamefont {Ahmed}},
  \bibinfo {author} {\bibfnamefont {V.~N.}\ \bibnamefont {Gamezo}}, \ and\
  \bibinfo {author} {\bibfnamefont {B.~D.}\ \bibnamefont {Taylor}},\ }\href
  {\doibase 10.1126/science.aau7365} {\bibfield  {journal} {\bibinfo  {journal}
  {Science}\ }\textbf {\bibinfo {volume} {366}},\ \bibinfo {pages} {eaau7365}
  (\bibinfo {year} {2019})}\BibitemShut {NoStop}%
\bibitem [{\citenamefont {Mannucci}\ \emph {et~al.}(2006)\citenamefont
  {Mannucci}, \citenamefont {Della~Valle},\ and\ \citenamefont
  {Panagia}}]{10.1111/j.1365-2966.2006.10501.x}%
  \BibitemOpen
  \bibfield  {author} {\bibinfo {author} {\bibfnamefont {F.}~\bibnamefont
  {Mannucci}}, \bibinfo {author} {\bibfnamefont {M.}~\bibnamefont
  {Della~Valle}}, \ and\ \bibinfo {author} {\bibfnamefont {N.}~\bibnamefont
  {Panagia}},\ }\href {\doibase 10.1111/j.1365-2966.2006.10501.x} {\bibfield
  {journal} {\bibinfo  {journal} {Monthly Notices of the Royal Astronomical
  Society}\ }\textbf {\bibinfo {volume} {370}},\ \bibinfo {pages} {773}
  (\bibinfo {year} {2006})},\ \Eprint
  {http://arxiv.org/abs/https://academic.oup.com/mnras/article-pdf/370/2/773/2903658/mnras0370-0773.pdf}
  {https://academic.oup.com/mnras/article-pdf/370/2/773/2903658/mnras0370-0773.pdf}
  \BibitemShut {NoStop}%
\bibitem [{\citenamefont {Matteucci}\ \emph {et~al.}(2006)\citenamefont
  {Matteucci}, \citenamefont {Panagia}, \citenamefont {Pipino}, \citenamefont
  {Mannucci}, \citenamefont {Recchi},\ and\ \citenamefont
  {Della~Valle}}]{10.1111/j.1365-2966.2006.10848.x}%
  \BibitemOpen
  \bibfield  {author} {\bibinfo {author} {\bibfnamefont {F.}~\bibnamefont
  {Matteucci}}, \bibinfo {author} {\bibfnamefont {N.}~\bibnamefont {Panagia}},
  \bibinfo {author} {\bibfnamefont {A.}~\bibnamefont {Pipino}}, \bibinfo
  {author} {\bibfnamefont {F.}~\bibnamefont {Mannucci}}, \bibinfo {author}
  {\bibfnamefont {S.}~\bibnamefont {Recchi}}, \ and\ \bibinfo {author}
  {\bibfnamefont {M.}~\bibnamefont {Della~Valle}},\ }\href {\doibase
  10.1111/j.1365-2966.2006.10848.x} {\bibfield  {journal} {\bibinfo  {journal}
  {Monthly Notices of the Royal Astronomical Society}\ }\textbf {\bibinfo
  {volume} {372}},\ \bibinfo {pages} {265} (\bibinfo {year} {2006})},\ \Eprint
  {http://arxiv.org/abs/https://academic.oup.com/mnras/article-pdf/372/1/265/5782472/mnras0372-0265.pdf}
  {https://academic.oup.com/mnras/article-pdf/372/1/265/5782472/mnras0372-0265.pdf}
  \BibitemShut {NoStop}%
\bibitem [{\citenamefont {Maoz}\ \emph {et~al.}(2014)\citenamefont {Maoz},
  \citenamefont {Mannucci},\ and\ \citenamefont
  {Nelemans}}]{doi:10.1146/annurev-astro-082812-141031}%
  \BibitemOpen
  \bibfield  {author} {\bibinfo {author} {\bibfnamefont {D.}~\bibnamefont
  {Maoz}}, \bibinfo {author} {\bibfnamefont {F.}~\bibnamefont {Mannucci}}, \
  and\ \bibinfo {author} {\bibfnamefont {G.}~\bibnamefont {Nelemans}},\ }\href
  {\doibase 10.1146/annurev-astro-082812-141031} {\bibfield  {journal}
  {\bibinfo  {journal} {Annual Review of Astronomy and Astrophysics}\ }\textbf
  {\bibinfo {volume} {52}},\ \bibinfo {pages} {107} (\bibinfo {year} {2014})},\
  \Eprint
  {http://arxiv.org/abs/https://doi.org/10.1146/annurev-astro-082812-141031}
  {https://doi.org/10.1146/annurev-astro-082812-141031} \BibitemShut {NoStop}%
\end{thebibliography}

\begin{thebibliography}{3}%
\makeatletter
\providecommand \@ifxundefined [1]{%
 \@ifx{#1\undefined}
}%
\providecommand \@ifnum [1]{%
 \ifnum #1\expandafter \@firstoftwo
 \else \expandafter \@secondoftwo
 \fi
}%
\providecommand \@ifx [1]{%
 \ifx #1\expandafter \@firstoftwo
 \else \expandafter \@secondoftwo
 \fi
}%
\providecommand \natexlab [1]{#1}%
\providecommand \enquote  [1]{``#1''}%
\providecommand \bibnamefont  [1]{#1}%
\providecommand \bibfnamefont [1]{#1}%
\providecommand \citenamefont [1]{#1}%
\providecommand \href@noop [0]{\@secondoftwo}%
\providecommand \href [0]{\begingroup \@sanitize@url \@href}%
\providecommand \@href[1]{\@@startlink{#1}\@@href}%
\providecommand \@@href[1]{\endgroup#1\@@endlink}%
\providecommand \@sanitize@url [0]{\catcode `\\12\catcode `\$12\catcode
  `\&12\catcode `\#12\catcode `\^12\catcode `\_12\catcode `\%12\relax}%
\providecommand \@@startlink[1]{}%
\providecommand \@@endlink[0]{}%
\providecommand \url  [0]{\begingroup\@sanitize@url \@url }%
\providecommand \@url [1]{\endgroup\@href {#1}{\urlprefix }}%
\providecommand \urlprefix  [0]{URL }%
\providecommand \Eprint [0]{\href }%
\providecommand \doibase [0]{http://dx.doi.org/}%
\providecommand \selectlanguage [0]{\@gobble}%
\providecommand \bibinfo  [0]{\@secondoftwo}%
\providecommand \bibfield  [0]{\@secondoftwo}%
\providecommand \translation [1]{[#1]}%
\providecommand \BibitemOpen [0]{}%
\providecommand \bibitemStop [0]{}%
\providecommand \bibitemNoStop [0]{.\EOS\space}%
\providecommand \EOS [0]{\spacefactor3000\relax}%
\providecommand \BibitemShut  [1]{\csname bibitem#1\endcsname}%
\let\auto@bib@innerbib\@empty
\bibitem [{\citenamefont {Caplan}\ and\ \citenamefont
  {Horowitz}(2017)}]{CaplanRMP}%
  \BibitemOpen
  \bibfield  {author} {\bibinfo {author} {\bibfnamefont {M.~E.}\ \bibnamefont
  {Caplan}}\ and\ \bibinfo {author} {\bibfnamefont {C.~J.}\ \bibnamefont
  {Horowitz}},\ }\href {\doibase 10.1103/RevModPhys.89.041002} {\bibfield
  {journal} {\bibinfo  {journal} {Rev. Mod. Phys.}\ }\textbf {\bibinfo {volume}
  {89}},\ \bibinfo {pages} {041002} (\bibinfo {year} {2017})}\BibitemShut
  {NoStop}%
\bibitem [{\citenamefont {Engstrom}\ \emph {et~al.}(2016)\citenamefont
  {Engstrom}, \citenamefont {Yoder},\ and\ \citenamefont
  {Crespi}}]{engstrom2016crystal}%
  \BibitemOpen
  \bibfield  {author} {\bibinfo {author} {\bibfnamefont {T.}~\bibnamefont
  {Engstrom}}, \bibinfo {author} {\bibfnamefont {N.}~\bibnamefont {Yoder}}, \
  and\ \bibinfo {author} {\bibfnamefont {V.}~\bibnamefont {Crespi}},\
  }\href@noop {} {\bibfield  {journal} {\bibinfo  {journal} {The Astrophysical
  Journal}\ }\textbf {\bibinfo {volume} {818}},\ \bibinfo {pages} {183}
  (\bibinfo {year} {2016})}\BibitemShut {NoStop}%
\bibitem [{\citenamefont {Caplan}(2020)}]{Caplan2020}%
  \BibitemOpen
  \bibfield  {author} {\bibinfo {author} {\bibfnamefont {M.~E.}\ \bibnamefont
  {Caplan}},\ }\href {\doibase 10.1103/PhysRevE.101.023201} {\bibfield
  {journal} {\bibinfo  {journal} {Phys. Rev. E}\ }\textbf {\bibinfo {volume}
  {101}},\ \bibinfo {pages} {023201} (\bibinfo {year} {2020})}\BibitemShut
  {NoStop}%
\end{thebibliography}

\providecommand{\noopsort}[1]{}\providecommand{\singleletter}[1]{#1}%

\end{document}